\documentclass[letterpaper,preprintnumbers,floats,floatfix,showpacs,amssymb,prd,twocolumn,superscriptaddress,nofootinbib]{revtex4}
\usepackage{amssymb,amsmath}
\usepackage{epsfig,hyperref}
\usepackage[svgnames]{xcolor}
\usepackage{pgf,tikz}
\usepackage{epstopdf}
\usepackage[british]{babel}

\makeatletter
\DeclareRobustCommand*{\bfseries}{%
  \not@math@alphabet\bfseries\mathbf
  \fontseries\bfdefault\selectfont
  \boldmath
}
\makeatother

\usepackage{geometry}
\def\be{\begin{equation}}
\def\ee{\end{equation}}
\def\beq{\begin{eqnarray}}
\def\eeq{\end{eqnarray}}

\makeatletter

\newcommand{\arXiv}[2][]{\href{http://arxiv.org/abs/#2}{\texttt{arXiv:#2\@ifempty{#1}{}{ [#1]}}}}
\makeatother

\begin{document}

\title{Spherical collapse in $f(R)$ gravity and the Belinskii-Khalatnikov-Lifshitz conjecture}

\author{Jun-Qi Guo}%
\email{jga35@sfu.ca}
\affiliation{
Department of Physics, Simon Fraser University\\
8888 University Drive, Burnaby, British Columbia V5A 1S6, Canada
}
\author{Daoyan Wang}
\email{dwang@phas.ubc.ca}
\affiliation{
Department of Physics \& Astronomy, University of British Columbia\\
6224 Agricultural Road, Vancouver, British Columbia V6T 1Z1, Canada
}
\author{Andrei V. Frolov}
\email{frolov@sfu.ca}
\affiliation{
Department of Physics, Simon Fraser University\\
8888 University Drive, Burnaby, British Columbia V5A 1S6, Canada
}

\date{\today}

\begin{abstract}
Spherical scalar collapse in $f(R)$ gravity is studied numerically in double-null coordinates in the Einstein frame. Dynamics in the vicinity of the singularity of the formed black hole is examined via mesh refinement and asymptotic analysis. Before the collapse, the scalar degree of freedom $f'$ is coupled to a physical scalar field, and general relativity is restored. During the collapse, the major energy of the physical scalar field moves to the center. As a result, $f'$ loses the coupling and becomes light, and gravity transits from general relativity to $f(R)$ gravity.
Due to strong gravity from the singularity and the low mass of $f'$, $f'$ will cross the minimum of the potential and approach zero. Therefore, the dynamical solution is significantly different from the static solution of the black hole in $f(R)$ gravity---it is not the de Sitter-Schwarzschild solution as one might have expected. $f'$ tries to suppress the evolution of the physical scalar field, which is a dark energy effect.
As the singularity is approached, metric terms are dominant over other terms. The Kasner solution for spherical scalar collapse in $f(R)$ theory is obtained and confirmed by numerical results. These results support the Belinskii-Khalatnikov-Lifshitz conjecture well.

\end{abstract}
\maketitle

\vspace{10pt}

\section{Introduction}
General relativity is a milestone in gravitation. However, some problems in general relativity, e.g., the nonrenormalizability of general relativity and the singularity problems in black hole physics and in the early Universe, imply that general relativity may not be the final gravitational theory~\cite{Stelle,Starobinsky_1,Biswas_1,Biswas_2}. Theoretical and observational explorations in cosmology and astrophysics, e.g., inflation, the orbital velocities of galaxies in clusters and the cosmic acceleration, also encourage considerations of new gravitational theories~\cite{Brans_1961,Starobinsky_1,Milgrom,Damour,Carroll,Hu_0705,Starobinsky_2}. Among various modified gravity theories, $f(R)$ gravity is a natural extension of general relativity. In this theory, the Ricci scalar, $R$, in the Einstein-Hilbert action is replaced by an arbitrary function of the Ricci scalar,
\be S=\frac{1}{16\pi G}\int d^{4}x \sqrt{-g}f(R) + S_{m}, \label{f_R_action} \ee
where $G$ is the Newtonian constant, and $S_{m}$ is the matter term in the action~\cite{Carroll, Starobinsky_2, Hu_0705}. The models of this type became popular in cosmology, with people trying to attribute the late-time accelerated expansion of the Universe to gravitational degrees of freedom. (See Refs.~\cite{Sotiriou_1,Tsujikawa1} for reviews of $f(R)$ theory.)

Black hole physics and spherical collapse are important platforms for understanding gravity. (For reviews of gravitational collapse and spacetime singularities, see Refs.~\cite{Berger_2002,Joshi_2007,Henneaux,Kamenshchik,Joshi_2011,Belinski_1404}.) Historically, some static solutions for black holes have been obtained analytically. The \lq\lq no-hair\rq\rq~theorem states that a stationary black hole can be described by only a few parameters~\cite{Wheeler}. Hawking showed that stationary black holes as the final states of Brans-Dicke collapses are identical to those in general relativity~\cite{Hawking}. In Ref.~\cite{Bekenstein}, a novel \lq\lq no-hair\rq\rq~theorem was proven. In this theorem, the scalar field, surrounding an asymptotically flat, static, spherically symmetric black hole, is assumed to be minimally coupled to gravity, and to have a non-negative energy density. In this case, the black hole must be a Schwarzschild black hole. This result is also valid if the scalar field has a potential whose global minimum is zero. The possible black hole solutions were explored in scalar-tensor gravity, including $f(R)$ gravity, by Sotiriou and Faraoni. If black holes were to be isolated from the cosmological background, they would have a Schwarzschild
solution~\cite{Sotiriou_2}.

As astrophysical black holes are expected to come from collapses of matter, studying collapse processes, especially spherical collapses, is an instructive way to explore black hole physics and to verify the results on stationary black holes as well. The Oppenheimer-Snyder solution provides an analytic description of the spherical dust collapse into a Schwarzschild black hole~\cite{Oppenheimer}. The Lema\^{\i}tre-–Tolman--Bondi solution describes a spherically symmetric inhomogeneous universe filled with dust matter~\cite{Lemaitre,Tolman,Bondi}. However, due to the nonlinearity of Einstein field equations, in most other cases, the collapse solutions have to be searched for numerically. Simulations of spherical collapse in Brans-Dicke theory were implemented in Refs.~\cite{Scheel_1,Scheel_2,Shibata}, confirming Hawking's conclusion that stationary black holes as the final states of Brans-Dicke collapses are identical to those in general relativity~\cite{Hawking}. In Ref.~\cite{Hertog}, numerical integration of the Einstein equations outwards from the horizon was performed. The results strongly supported the new \lq\lq no-hair\rq\rq~theorem presented in Ref.~\cite{Bekenstein}. Recently, the dynamics of single and binary black holes in scalar-tensor theories in the presence of a scalar field was studied in Ref.~\cite{Berti}, in which the potential for scalar-tensor theories is set to zero and the source scalar field is assumed to have a constant gradient.

Although $f(R)$ theory is equivalent to scalar-tensor theories, it is of a unique type. In $f(R)$ theory, the potential is related to the function $f(R)$ or the Ricci scalar $R$ by $V'(\chi)\equiv(2f-\chi R)/3$, with $\chi\equiv f'$. In dark-energy-oriented $f(R)$ gravity, the de Sitter curvature obtained from $V'(\chi)\equiv(2f-\chi R)/3=0$ is expected to drive the cosmic acceleration. Consequently, the minimum of the potential cannot be zero. Therefore, spherical collapse in $f(R)$ theory has rich phenomenology and is worth exploring in depth, although some studies have been implemented in scalar-tensor theories. In Ref.~\cite{Cembranos}, the gravitational collapse of a uniform dust cloud in $f(R)$ gravity was analyzed; the scale factor and the collapsing time were computed. In Ref.~\cite{Senovilla}, the junction conditions through the hypersurface separating the exterior and the interior of the global gravitational field
in $f(R)$ theory were derived. In Ref.~\cite{Hwang_1110}, a charged black hole from gravitational collapse in $f(R)$ gravity was obtained. However, to a large extent, a general collapse in scalar-tensor theories [especially in $f(R)$ theory], in which the global minimum of the potential is nonzero, still remains unexplored as of yet.
In addition to black hole formation, large-scale structure is another formation that can be modeled. In Refs.~\cite{Kopp,Barreira}, with the scalar fields assumed to be quasistatic, simulations of dark matter halo formation were implemented in $f(R)$ gravity and Galileon gravity, respectively.

Another motivation comes from the study of the dynamics as one approaches the singularity. The Belinskii, Khalatnikov, and Lifshitz (BKL) conjecture states  that as the singularity is approached, the dynamical terms will dominate the spatial terms in the Einstein field equations, the metric terms will dominate the matter field terms, and the metric components and the matter fields are described by the Kasner solution~\cite{Belinskii,Landau,Kasner,Wainwright}. The BKL conjecture was verified numerically for the singularity formation in a closed cosmology in Refs.~\cite{Berger,Garfinkel_1}. It was also confirmed in Ref.~\cite{Garfinkel_2} with a test scalar field approaching the singularity of a black hole, whose metric is described by a spatially flat dust Friedmann$–-$Lema\^{\i}tre$–-$Robertson$–-$Walker spacetime. In Ref.~\cite{Ashtekar}, the BKL conjecture in the Hamiltonian framework was examined, in an attempt to understand the implications of the BKL conjecture for loop quantum gravity. In this paper, we consider a scalar field collapse in $f(R)$ gravity. We study the evolution of the spacetime, the physical scalar field $\psi$, and the scalar degree of freedom $f'$ throughout the whole collapse process and also in the vicinity of the singularity of the formed black hole.

Regarding simulations of gravitational collapses and binary black holes in gravitational theories beyond general relativity, in addition to the references mentioned above, in Ref.~\cite{Harada}, the generation and propagation of the scalar gravitational wave from a spherically symmetric and homogeneous dust collapse in scalar-tensor theories were computed numerically, with the backreaction of the scalar wave on the spacetime being neglected. Scalar gravitational waves generated from stellar radial oscillations in scalar-tensor theories were computed in Ref.~\cite{Sotani}. The response of the Brans-Dicke field during gravitational collapse was studied in Ref.~\cite{Hwang_2010}. Charge collapses in dilaton gravity were explored in Refs.~\cite{Frolov_0504,Frolov_0604,Borkowska}. Binary black hole mergers in $f(R)$ theory were simulated in Ref.~\cite{Cao}.

A viable dark energy $f(R)$ model should be stable~\cite{Nunez,Dolgov}, able to generate a cosmological evolution consistent with the observations~\cite{Amendola,Guo_1305}, and able to pass the solar system tests~\cite{Hu_0705,Justin1,Justin2,Chiba,Tamaki_0808,Tsujikawa_0901,Guo_1306}. These requirements imply that this model should be reduced to general relativity at high curvature scale, $R\gg\Lambda$, and mainly modifies general relativity at low curvature scale, $R\sim \Lambda$, where $\Lambda$ is the currently observed effective cosmological constant. We take two typical viable $f(R)$ models, the Hu-Sawicki model~\cite{Hu_0705} and the Starobinsky model~\cite{Starobinsky_2}, as sample $f(R)$ models. We perform the simulations in the double-null coordinates proposed by Christodoulou~\cite{Christodoulou}. These coordinates have been used widely, because they have the horizon-penetration advantage and also allow us to study the global structure of spacetime~\cite{Frolov_2004,Frolov_0504,Frolov_0604,Sorkin,Hwang_2010,Borkowska,Hwang_1105,Golod}. The results show that a black hole can be formed. During the collapse, the scalar field $f'$ is decoupled from the matter density and becomes light. Simultaneously, the Ricci scalar decreases, and the modification term in the function $f(R)$ becomes important. The lightness of $f'$ and the gravity from the scalar sphere, which forms a black hole later, make the scalar field $f'$ cross the minimum of the potential (also called a de Sitter point), and then approach zero near the singularity. The asymptotic expressions for the metric components and scalar fields are obtained. They are the Kasner solution. These results support the BKL conjecture.

To a large extent, the features of $f(R)$ theory are defined by the shape of the potential. Local tests and cosmological dynamics of $f(R)$ theory are closely related to the right side and the minimum area of the potential~\cite{Justin1,Justin2,Guo_1305,Guo_1306}. In the early Universe, the scalar degree of freedom $f'$ is coupled to the matter density. In the later evolution, $f'$ is decoupled from the matter density and goes down toward the minimum of the potential, and eventually stops at the minimum after some oscillations. Interestingly, studies of collapses draw one's attention to the left side of the potential.

The paper is organized as follows.
In Sec.~\ref{sec:framework}, we introduce the framework of the collapse, including the formalism of $f(R)$ theory, double-null coordinates, and the Hu-Sawicki model.
In Sec.~\ref{sec:set_up}, we set up the numerical structure, including discretizing the equations of motion, defining initial and boundary conditions, and implementing the numerical tests.
In Sec.~\ref{sec:results}, numerical results are presented.
In Sec.~\ref{sec:view_JF}, we discuss numerical results from the point of view of the Jordan frame.
In Sec.~\ref{sec:collapse_general}, we consider collapses in more general models.
Section~\ref{sec:conclusions} summarizes our work.

\section{Framework\label{sec:framework}}
In this section, we build the framework of spherical scalar collapse in $f(R)$ theory. To utilize the developed tools in numerical relativity, $f(R)$ gravity is transformed from the Jordan frame into the Einstein frame. In order to study the global structure of the spacetime, and the dynamics of the spacetime and the source fields near the singularity, we simulate the collapse in double-null coordinates. A typical $f(R)$ model, the Hu-Sawicki model, is chosen as an example model. This paper gives the first detailed results on numerical simulations of fully dynamical spherical collapse in $f(R)$ gravity toward a black hole formation.

\subsection{$f(R)$ theory}

The equivalent of the Einstein equation in $f(R)$ gravity reads
\be f'R_{\mu\nu}-\frac{1}{2}f g_{\mu\nu} -\left(\nabla_{\mu}\nabla_{\nu}-g_{\mu\nu} \Box\right) f'
= 8\pi GT_{\mu\nu},\label{gravi_eq_fR} \ee
where $f'$ denotes the derivative of the function $f$ with respect to its argument $R$, and $\Box$ is the usual
notation for the covariant D'Alembert operator $\Box\equiv\nabla_{\alpha}\nabla^{\alpha}$. The trace of Eq. (\ref{gravi_eq_fR}) is
\be \Box f'=\frac{1}{3}(2f-f'R) + \frac{8\pi G}{3}T, \label{trace_eq1}\ee
where $T$ is the trace of the stress-energy tensor $T_{\mu\nu}$. In general relativity, $f'\equiv1$ and $\Box f'\equiv0$. However, $\Box f'$ is generally not zero in $f(R)$ gravity. Therefore, compared to general relativity, there is a scalar degree of freedom, $f'$, in $f(R)$ gravity. Identifying $f'$ with a scalar degree of freedom by
\be \chi\equiv\frac{df}{dR}, \label{f_prime}\ee
and defining a potential $U(\chi)$ by
\be U'(\chi)\equiv\frac{dU}{d\chi}=\frac{1}{3}(2f-\chi R), \label{v_prime} \ee
one can rewrite Eq.~(\ref{trace_eq1}) as
\be \Box \chi=U'(\chi)+\frac{8\pi G}{3}T.\label{trace_eq2}\ee
In order to operate $f(R)$ gravity, it is instructive to cast the formulation of $f(R)$ gravity into a format similar to that of general relativity. We rewrite Eq.~(\ref{gravi_eq_fR}) as
\be G_{\mu \nu}=8\pi G \left[ T_{\mu \nu} + T_{\mu \nu} ^{(\text{eff})} \right],
\label{field_eq4} \ee
where

\begin{eqnarray}
8\pi GT_{\mu \nu}^{(\text{eff})} & = & \frac{f-f'R}{2}g_{\mu\nu}+\left(\nabla_{\mu}\nabla_{\nu}-g_{\mu\nu} \Box\right)f'\nonumber\\
 && + (1-f')G_{\mu\nu}.
 \label{tilde_T4}
\end{eqnarray}
$T^{\mu\nu}_{\text{(eff)}}$ is the energy-momentum tensor of the effective dark energy. It is guaranteed to be conserved, $T^{\mu\nu}_{\text{(eff)};\nu} = 0$.
Note that there are second-order derivatives of $f'$ in $T_{\mu\nu}^{\text{(eff)}}$. In order to make the formalism less complicated, we transform $f(R)$ gravity from the current frame, which is usually called the Jordan frame, into the Einstein frame. In the latter, the second-order derivatives of $f'$ are absent in the equations of motion for the metric components. The formalism can be treated as Einstein gravity coupled to two scalar fields. Therefore, we can use some results that have been developed in the numerical relativity community.

Rescaling $\chi$ by
\be
\kappa \phi \equiv \sqrt{\frac{3}{2}} \log \chi,
\label{rescale_f_prime}
\ee
one obtains the corresponding action of $f(R)$ gravity in the Einstein frame~\cite{Tsujikawa1}
\begin{eqnarray}
S_E & = & \int d^4 x\sqrt{-\tilde g}\left[\frac{1}{2\kappa^2} \tilde R -\frac{1}{2} \tilde g^{\mu\nu}\partial_\mu\phi\partial_\nu\phi-V(\phi)\right]\nonumber\\
 && + \int d^4 x\mathcal{L}_M(\tilde g_{\mu\nu} / {\chi(\phi)},\psi),
\end{eqnarray}
where $\kappa=\sqrt{8\pi G}$, $\tilde g_{\mu\nu}=\chi\cdot g_{\mu\nu}$, $V(\phi)\equiv (\chi R-f)/(2\kappa^2 \chi^2)$, $\psi$ is a matter field, and a tilde denotes that the quantities are in the Einstein frame. The Einstein field equations are
\be
\tilde G_{\mu\nu} = \kappa^2 \left[\tilde T_{\mu\nu}^{(\phi)}+\tilde T_{\mu\nu}^{(M)}\right],
\ee
where
\begin{align}
\tilde T_{\mu\nu}^{(\phi)} &= \partial_\mu\phi\partial_\nu\phi-\tilde g_{\mu\nu}\left[\frac{1}{2}\tilde g^{\alpha\beta}\partial_\alpha\phi\partial_\beta\phi+V(\phi)\right], \\
\tilde T_{\mu\nu}^{(M)} &= \frac{T_{\mu\nu}^{(M)}}{\chi}.
\end{align}
$T_{\mu\nu}^{(M)}$ is the ordinary energy-momentum tensor of the physical matter field in terms of $g_{\mu\nu}$ in the Jordan frame. We take a massless scalar field $\psi$ as the matter field for the collapse. Its energy-momentum tensor in the Einstein frame is
\begin{eqnarray}
  \hspace{-18pt}\tilde T_{\mu\nu}^{(M)}=\tilde T_{\mu\nu}^{(\psi)}
  &=&\frac{1}{\chi}\left(\partial_\mu \psi \partial_\nu \psi-\frac{1}{2}g_{\mu\nu}g^{\alpha\beta}\partial_\alpha\psi\partial_\beta \psi\right)\nonumber\\
  &=&\frac{1}{\chi} \left(\partial_\mu \psi \partial_\nu \psi - \frac{1}{2}\tilde g_{\mu\nu} \tilde g^{\alpha\beta}\partial_\alpha\psi\partial_\beta \psi\right),
\end{eqnarray}
which gives
\be \tilde T^{(M)}=\tilde T^{(\psi)} \equiv \tilde g^{\mu\nu} \tilde T_{\mu\nu}^{(\psi)}
= - \frac{\tilde g^{\alpha\beta}\partial_\alpha\psi\partial_\beta \psi}{\chi}.\nonumber\ee
The equations of motion for $\phi$ and $\psi$ can be derived from the Lagrange equations as
\be
\tilde \Box \phi - V'(\phi)+\kappa Q \tilde T^{(M)} = 0,
\label{fR:Box_phi}
\ee
\be
\tilde \Box \psi -\sqrt{\frac{2}{3}} ~\kappa \tilde g^{\mu\nu} \partial_\mu \phi\partial_\nu \psi = 0,
\label{fR:Box_psi}
\ee
where $Q \equiv -\chi_{,\phi}/(2\kappa \chi)=-1/\sqrt{6}$. In the Einstein frame, the potential for $\phi$ is written as
\be V(\phi)=\frac{\chi R-f}{2\kappa^2\chi ^2}. \label{potential_EF}\ee
Then we have
\be V'(\phi)=\frac{dV}{d\chi}\cdot \frac{d\chi}{d\phi}
=\frac{1}{\sqrt{6}}\frac{2f-\chi R}{\kappa \chi ^2}.\label{V_prime_EF}\ee

\subsection{Coordinate system}
We are interested in the singularity formation, the dynamics of the spacetime and the source fields near the singularity, and the global structure of the spacetime. The double-null coordinates described by Eq.~(\ref{double_null_metric}) are a viable choice to realize these objectives~\cite{Christodoulou}:
\begin{eqnarray}
ds^{2} & = & e^{-2\sigma}(-dt^2+dx^2)+r^2d\Omega^2 \nonumber\\
  & = & -4e^{-2\sigma}dudv+r^2d\Omega^2,
\label{double_null_metric}
\end{eqnarray}
where $\sigma$ and $r$ are functions of $(t,x)$, and $u[=(t-x)/2=\text{Const}]$ and $v[=(t + x)/2=\text{Const}]$ are outgoing and ingoing characteristics, respectively. The two-manifold metric
\be d\gamma^2=e^{-2\sigma}(-dt^2+dx^2)=-4e^{-2\sigma}dudv \nonumber\ee
is conformally flat. In these coordinates, one can know the speed of information propagation everywhere in advance.

The metric (\ref{double_null_metric}) is invariant for the rescaling $u\rightarrow U(u)$, $v\rightarrow V(v)$. We fix this gauge freedom by setting up initial and boundary conditions.

\subsection{The Hu-Sawicki model}
\begin{figure}
  \epsfig{file=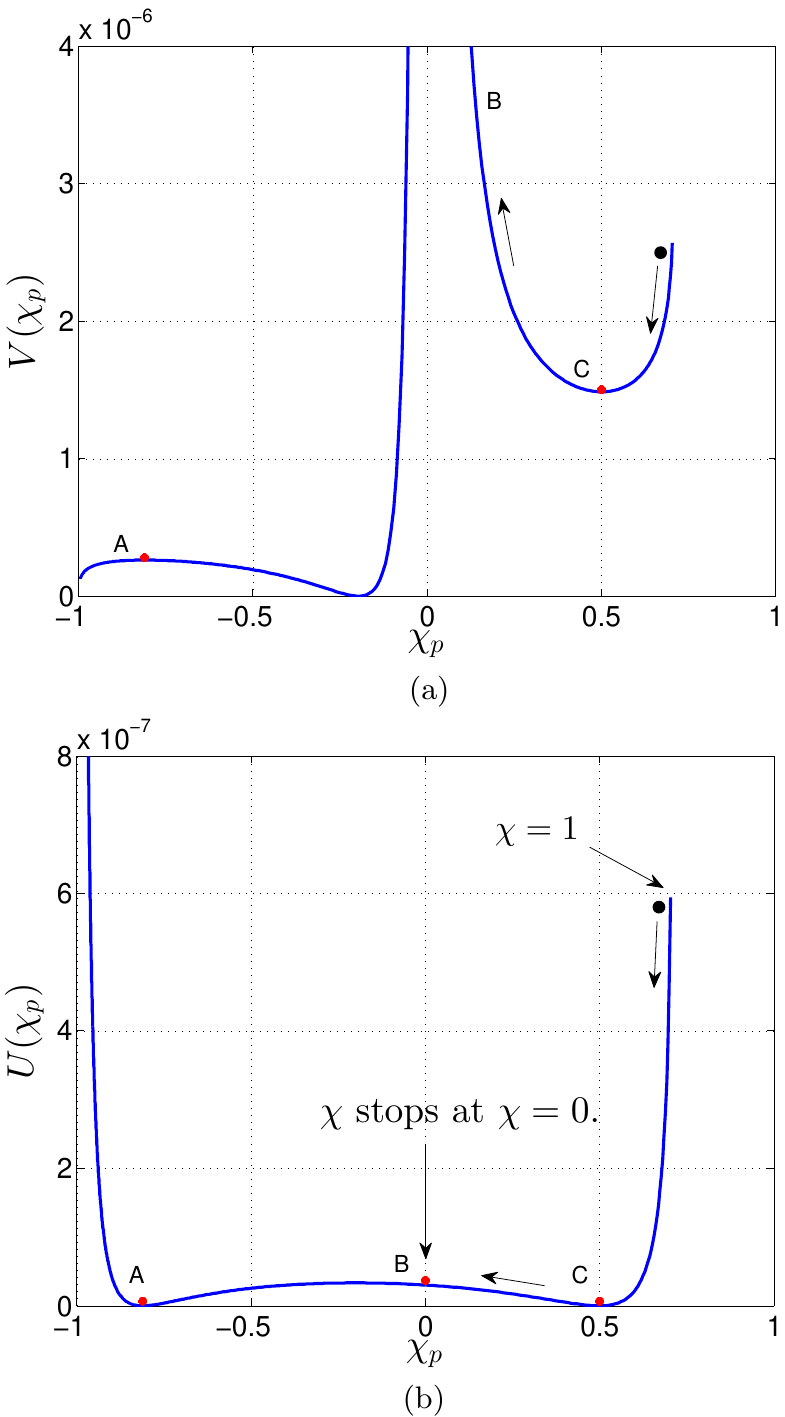, width=7.7cm,height=14.0cm}
  \caption{The potentials for the Hu-Sawicki model in (a) the Einstein frame and (b) the Jordan frame. $\chi_p$ is a compactified coordinate obtained via the $\text{Poincar\'e}$ transformation, $\chi_p=\chi/\sqrt{1+\chi^2}$, with $\chi\equiv f'$. The potential $V(\chi(\phi))$ in the Einstein frame is defined by Eq.~(\ref{V_Hu_Sawicki_EF}), and the potential $U(\chi)$ in the Jordan frame is defined by Eq.~(\ref{v_prime}). A scalar field can collapse to form a black hole. As the singularity of the black hole is approached, $\chi$ asymptotes to zero.}
  \label{fig:potential_Hu_Sawicki}
\end{figure}

For a viable dark energy $f(R)$ model, $f'$ has to be positive to avoid ghosts~\cite{Nunez}, and $f''$ has to be positive to avoid the Dolgov-Kawasaki instability~\cite{Dolgov}. The model should also be able to generate a cosmological evolution compatible with the observations~\cite{Amendola,Guo_1305} and to pass the solar system tests~\cite{Hu_0705,Justin1,Justin2,Chiba,Tamaki_0808,Tsujikawa_0901,Guo_1306}. Equivalently, general relativity should be restored at high curvature scale, $R\gg\Lambda$, and the $f(R)$ model mainly deviates from general relativity at low curvature scale, $R\sim \Lambda$. In this paper, we take the Hu-Sawicki $f(R)$ model as an example. This model reads~\cite{Hu_0705}
\be f(R)=R-R_{0}\frac{D_{1}R^n}{D_{2}R^n+R_{0}^n},\label{f_R_Hu_Sawicki_general}\ee
where $n$ is a positive parameter, $D_1$ and $D_2$ are dimensionless parameters, $R_{0}=8\pi G\bar{\rho}_0/3$, and $\bar{\rho}_0$ is the average matter density of the current Universe. We consider one of the simplest versions of this model, i.e., $n=1$,
\be f(R) = R-\frac{DR_{0} R}{R+R_{0}}, \label{f_R_Hu_Sawicki}\ee
where $D$ is a dimensionless parameter. In this model,
\be f'=1-\frac{DR_{0} ^2}{(R+R_{0})^2},\label{f_prime_Hu_Sawicki}\ee
\be R=R_{0} \left[\sqrt{\frac{D}{1-f'}}-1\right], \label{R_Hu_Sawicki}\ee
\be V(\phi)=\frac{DR_{0} R^2}{2\kappa^2{f'}^2(R+R_{0})^2},\label{V_Hu_Sawicki_EF}\ee
\be V'(\phi)=\frac{R^{3}}{\sqrt{6}\kappa f'^{2}(R+R_{0})^{2}}\left[1+(1-D)\frac{R_{0}}{R}\left(2+\frac{R_{0}}{R}\right) \right].
\label{V_prime_Hu_Sawicki}\ee
Equations (\ref{f_prime_Hu_Sawicki}) and (\ref{V_prime_Hu_Sawicki}) show that as long as the matter density is much greater than $R_{0}$, the curvature $R$ will trace the matter density well, $f'$ will be close to $1$ but not cross $1$, and general relativity will be restored. As implied in Eq.~(\ref{V_prime_Hu_Sawicki}), in order to make sure that the de Sitter curvature, for which $V'(\phi)=0$, has a positive value, the parameter $D$ needs to be greater than $1$. In this paper, we set $D$ and $R_{0}$ to $1.2$ and $5\times 10^{-6}$, respectively. Then, together with Eqs.~(\ref{V_Hu_Sawicki_EF}) and (\ref{V_prime_Hu_Sawicki}), these values imply that the radius of the de Sitter horizon is about $\sqrt{1/R_0}\sim 10^{3}$. Moreover, in the configuration of the initial conditions described in Sec.~\ref{sec:set_up_ic} and the above values of $D$ and $R_{0}$, the radius of the apparent horizon of the formed black hole is about $2.2$ [see Fig.~\ref{fig:apparent_horizon}.(b)]. The potential in the Einstein frame defined by Eq.~(\ref{V_Hu_Sawicki_EF}) and the potential in the Jordan frame defined by Eq.~(\ref{v_prime}) are plotted in Figs.~\ref{fig:potential_Hu_Sawicki}(a) and (b), respectively.

After explorations of spherical collapse for one of the simplest versions of the Hu-Sawicki model described by Eq.~(\ref{f_R_Hu_Sawicki}), we consider general cases. We let the parameters, $D$ in Eq.~(\ref{f_R_Hu_Sawicki}) and $n$ in Eq.~(\ref{f_R_Hu_Sawicki_general}), take different values. We also study spherical collapse for the Starobinsky model~\cite{Starobinsky_2}. All the results turn out to be similar.

\section{Numerical setup\label{sec:set_up}}
In this section, we present the numerical formalisms, including field equations, boundary conditions, initial conditions, discretization scheme, and numerical tests. The numerical code used in the paper is a generalized version of the one developed by one of the authors~\cite{Frolov_2004}.

\subsection{Field equations\label{sec:set_up_EoM}}
In this paper, we set $8\pi G$ to $1$. Details on components of Einstein tensor and energy-momentum tensor of a massive scalar field are given in Appendix~\ref{sec:appendix_A}. Then, in double-null coordinates (\ref{double_null_metric}), using
\be \tilde G^{t}_{t}+\tilde G^{x}_{x}
=\tilde T^{(\phi)t}_{\hphantom{ddd}t}+\tilde T^{(\phi)x}_{\hphantom{ddd}x}+\tilde T^{(\psi)t}_{\hphantom{ddd}t}+\tilde T^{(\psi)x}_{\hphantom{ddd}x},
\nonumber\ee
one obtains the equation of motion for the metric component $r$,
\be r(-r_{,tt}+r_{,xx}) + (-r_{,t}^2+r_{,x}^2) = e^{-2\sigma}(1-r^2V), \label{equation_r}\ee
where $r_{,t}\equiv dr/dt$, and other quantities are defined analogously. Equation (\ref{equation_r}) involves a delicate cancellation of terms at both small and large $r$, which makes it susceptible to discretization errors. In order to avoid this problem, when $r$ is not too large, we define $\eta\equiv r^2$, and integrate the equation of motion for $\eta$ instead. The equation of motion for $\eta$ can be obtained by rewriting Eq.~(\ref{equation_r}) as~\cite{Frolov_2004}
\be -\eta_{,tt}+\eta_{,xx}=2e^{-2\sigma}\left(1-r^2V\right).\label{equation_r_2}\ee
When $r$ is very large, the delicate cancellation problem can be avoided by using a new variable $\rho\equiv1/r$, instead~\cite{Frolov_2004}.
$\tilde G^{\theta}_{\theta} =\tilde T^{(\phi)\theta}_{\hphantom{ddd}\theta}+\tilde T^{(\psi)\theta}_{\hphantom{ddd}\theta}$ provides the equation of motion for $\sigma$,
\be
\begin{split}
&-\sigma_{,tt}+\sigma_{,xx} - \frac{-r_{,tt}+r_{,xx}}{r} - \frac{1}{2}(-\phi_{,t}^2+\phi_{,x}^2) \\
&- \frac{1}{2\chi}(-\psi_{,t}^2+\psi_{,x}^2) = e^{-2\sigma}V.
\end{split}
\label{equation_sigma}
\ee

In double-null coordinates, the dynamical equations for $\phi$ (\ref{fR:Box_phi}) and $\psi$ (\ref{fR:Box_psi}) become, respectively,
\be
\begin{split}
&(-\phi_{,tt}+\phi_{,xx})+\frac{2}{r}(-r_{,t}\phi_{,t}+r_{,x}\phi_{,x})\\
&=e^{-2\sigma}\left[V'(\phi)+\frac{1}{\sqrt{6}}\kappa \tilde T^{(\psi)}\right],
\end{split}
\label{equation_phi}
\ee
\be
\begin{split}
&(-\psi_{,tt}+\psi_{,xx})+\frac{2}{r}(-r_{,t}\psi_{,t}+r_{,x}\psi_{,x})\\
&=\sqrt{\frac{2}{3}}~\kappa (-\phi_{,t}\psi_{,t}+\phi_{,x}\psi_{,x}),
\end{split}
\label{equation_psi}
\ee
where
\be \tilde T^{(\psi)}=\frac{e^{2\sigma}(\psi_{,t}^2-\psi_{,x}^2)}{\chi}.\label{T_psi}\ee

The $\{uu\}$ and $\{vv\}$ components of the Einstein equations yield the constraint equations:
\be r_{,uu}+2\sigma_{,u}r_{,u}=-\frac{r}{2}\left(\phi_{,u}^2+\frac{\psi_{,u}^2}{\chi}\right), \label{constraint_eq_uu}\ee
\be r_{,vv}+2\sigma_{,v}r_{,v}=-\frac{r}{2}\left(\phi_{,v}^2+\frac{\psi_{,v}^2}{\chi}\right). \label{constraint_eq_vv}\ee
Via the definitions of $u=(t-x)/2$ and $v=(t+x)/2$, the constraint equations can be expressed in $(t,x)$ coordinates.
Equations $(\ref{constraint_eq_vv})-(\ref{constraint_eq_uu})$ and $(\ref{constraint_eq_vv})+(\ref{constraint_eq_uu})$ generate the constraint equations
for $\{tx\}$ and $\{tt\}+\{xx\}$ components, respectively,
\be r_{,tx}+r_{,t}\sigma_{,x}+r_{,x}\sigma_{,t}+\frac{r}{2}\phi_{,t}\phi_{,x}+\frac{r}{2\chi}\psi_{,t}\psi_{,x}=0,
\label{constraint_eq_xt}
\ee
\be
\begin{split}
&r_{,tt}+r_{,xx}+2r_{,t}\sigma_{,t}+2r_{,x}\sigma_{,x}+\frac{r}{2}(\phi_{,t}^2+\phi_{,x}^2) \\
&+ \frac{r}{2\chi}(\psi_{,t}^2+\psi_{,x}^2)=0.
\end{split}
\label{constraint_eq_xx_tt}
\ee

Regarding the equation of motion for $\sigma$, the term $(-r_{,tt}+r_{,xx})/r$ in Eq.~(\ref{equation_sigma}) can create big errors near the center $x=r=0$. To circumvent this problem, we use the constraint equation (\ref{constraint_eq_uu}) alternatively~\cite{Frolov_2004}. A new variable $g$ is defined as
\be g=-2\sigma-\log(-r_{,u}). \label{g_definition}\ee
Then, Eq.~(\ref{constraint_eq_uu}) can be written as the equation of motion for $g$,
\be g_{,u}=\frac{r}{2}\frac{\phi_{,u}^2 + \psi_{,u}^2/\chi }{r_{,u}}. \label{equation_g}\ee
In the numerical integration, once the values of $g$ and $r$ at the advanced level are obtained, the value of $\sigma$ at the current level will be computed from Eq.~(\ref{g_definition}).

\subsection{Boundary conditions}
In this paper, the range for the spatial coordinate is $x\in[0 \mbox{ } 22]$. The value of $22$ is chosen such that it is much less than the radius of the de Sitter horizon $(\sim \sqrt{1/R_0}\sim 10^{3})$, and it is much greater than the dynamical scale.

At the inner boundary where $x=0$, $r$ is always set to zero. The terms, $2(-r_{,t}\phi_{,t}+r_{,x}\phi_{,x})/r$ in Eq.~(\ref{equation_phi}) and $2(-r_{,t}\psi_{,t}+r_{,x}\psi_{,x})/r$ in Eq.~(\ref{equation_psi}), need to be regular at $r=x=0$. Since $r$ is always set to zero at the center, so is $r_{,t}$. Therefore, we enforce $\phi$ and $\psi$ to satisfy the following conditions:
\be \phi_{,x}=0, \hphantom{ddd} \psi_{,x}=0.\nonumber\ee
The boundary condition for $g$ at $r=0$ is obtained via extrapolation.

Considering the outer boundary, since one cannot include infinity on the grid, one needs to put a cutoff at $x$, where the radius $r$ is set to a constant. In this paper, we are mainly interested in the dynamics around the horizon and the dynamics near the singularity of the formed black hole. Dynamics in these regions will not be affected by the outer boundary conditions, as long as the spatial range of $x$ is large enough compared to the time range needed for black hole formation. In this paper, we set up the outer boundary conditions via extrapolation.

\subsection{Initial conditions\label{sec:set_up_ic}}
For any dynamical system whose evolution is governed by a second-order time derivative equation, its evolution is uniquely determined by setting the values of the dynamical variable and its first-order time derivative at any given instant. We set the initial data to be time-symmetric as follows:
\be r_{,t}=\sigma_{,t}=\phi_{,t}=\psi_{,t}=0 \hphantom{ddd} \mbox{at} \hphantom{d} t=0. \label{ic_t_0}\ee
In this case, the constraint equation (\ref{constraint_eq_xt}) is satisfied identically.

We set the initial value of $\psi(r)$ at $t=0$ as
\be \psi(r)=Q\cdot\tanh\left[(r-r_0)^2\right],\label{psi_ic}\ee
where $Q$ and $r_0$ take the values of $0.5$ and $5$, respectively. The initial value of $\phi(r)$ can be arbitrary as long as it is negative. [See Eq.~(\ref{rescale_f_prime}) and note that $\chi\equiv f'<1$.] Here we choose its value as that in a static system and weak-field limit $r=x$ and $\sigma=0$. In this case, the equation of motion for $\phi$, Eq.~(\ref{equation_phi}), becomes
\be \frac{d^2\phi}{dr^2}+\frac{2}{r}\frac{d\phi}{dr}=V'(\phi)+\frac{1}{\sqrt{6}}\kappa \tilde T^{(\psi)}.\ee
We solve this equation for initial $\phi(r)$ with Newton's iteration method, enforcing $d\phi/dr=0$ at $r=0$ and $\phi$ to stay at the minimum of the potential $V(\phi)$ at the outer boundary. [Note that $\tilde T^{(\psi)}=0$ at the outer boundary.]

We define a local mass by
\be p\equiv\tilde g^{\mu\nu}r_{,\mu}r_{,\nu}=1-\frac{2m}{r}. \label{define_local_mass}\ee
Using $r_{,t}=0$, $x=v-u$, and $t=u+v$, we have $r_u=-r_x$ at $t=0$. Then, in the double-null coordinates described by (\ref{double_null_metric}), Eq.~(\ref{define_local_mass}) implies that
\be e^{-2\sigma}=\frac{r_{,x}^2}{p}. \label{e_2sigma_ic}\ee
On the other hand, from Eq.~(\ref{g_definition}), one obtains at $t=0$,
\be r_{,u}=-r_{,x}=-e^{-2\sigma}e^{-g}. \label{r_u_ic}\ee
The combination of Eqs.~(\ref{e_2sigma_ic}) and (\ref{r_u_ic}) provides the equation for $r$,
\be r_{,x}=\left(1-\frac{2m}{r}\right)e^g. \label{r_x_ic}\ee

In addition to assigning $r_{,t}=\phi_{,t}=0$ at $t=0$, we set $r_{,tt}=0$ at $t=0$ to fix the gauge freedom. Consequently, Eqs.~(\ref{constraint_eq_xx_tt}) and (\ref{equation_r}) become, respectively,
\be r_{,xx}+2r_{,x}\sigma_{,x}=-\frac{r}{2}\left(\phi_{,x}^2 + \frac{\psi_{,x}^2}{\chi}\right), \label{constraint_eq_ic}\ee
\be e^{2\sigma}r_{,xx}=-rV+\frac{2m}{r^2}.\label{equation_r_ic}\ee
Differentiating Eq.~(\ref{e_2sigma_ic}) with respect to $r$ yields
\be e^{2\sigma}(2\sigma_{,x}r_{,x}+2r_{,xx})=-\frac{2m_{,r}}{r}+\frac{2m}{r^2}.\label{e_2sigma_diff}\ee
Substituting Eqs.~(\ref{constraint_eq_ic}) and (\ref{equation_r_ic}) into (\ref{e_2sigma_diff}) generates the equation for $m$
\be m_{,r}=\frac{r^2}{2}\left[V+\frac{1}{2}e^{2\sigma}\left(\phi_{,x}^2 + \frac{\psi_{,x}^2}{\chi}\right)\right]. \label{m_r_ic}\ee
Moreover, with Eqs.~(\ref{define_local_mass}) and (\ref{e_2sigma_ic}), we have
\begin{eqnarray}
e^{2\sigma}\left(\phi_{,x}^2 + \frac{\psi_{,x}^2}{\chi}\right) & = & e^{2\sigma}\left(\phi_{,r}^2 + \frac{\psi_{,r}^2}{\chi}\right)r_{,x}^2\nonumber\\
 & = & \left(\phi_{,r}^2 + \frac{\psi_{,r}^2}{\chi}\right)\left(1-\frac{2m}{r}\right).
\nonumber
\end{eqnarray}
Then, Eq.~(\ref{m_r_ic}) can be rewritten as
\be
m_{,r}=\frac{r^2}{2}\left[V+\frac{1}{2}\left(1-\frac{2m}{r}\right)\left(\phi_{,r}^2 + \frac{\psi_{,r}^2}{\chi}\right)\right].
\label{m_r_ic2}
\ee

The equation for $g$ at $t=0$ can be obtained from Eq.~(\ref{equation_g}),
\be g_{,r}=\frac{r}{2}\left(\phi_{,r}^2 + \frac{\psi_{,r}^2}{\chi}\right). \label{g_r_ic}\ee
We obtain the initial values of $r$, $m$, and $g$ at $t=0$ by integrating Eqs.~(\ref{r_x_ic}), (\ref{m_r_ic2}), and (\ref{g_r_ic}) via the fourth-order Runge-Kutta method. The values of $r$, $\sigma$, $f'(=\exp{\sqrt{2/3}\kappa \phi})$, and $\psi$ at $t=0$ are plotted in Fig.~\ref{fig:Evolution_metric_fields}.

In this paper, we implement a leapfrog scheme, which is a three-level scheme and requires initial data on two different time levels. With the initial data at $t=0$, we compute the data at $t=\Delta t$ with a second-order Taylor series expansion. Take the variable $\phi$ as an example:
\be \phi|_{t=\Delta t}=\phi|_{t=0}+\phi_{,t}|_{t=0}\Delta t +\frac{1}{2}\phi_{,tt}|_{t=0}(\Delta t)^2.\label{ic_taylor_expansion}\ee
The values of $\phi|_{t=0}$ and $\phi_{,t}|_{t=0}$ are set up as discussed above, and the value of $\phi_{,tt}|_{t=0}$ can be obtained from the equation of motion for $\phi$ (\ref{equation_phi}).

\subsection{Discretization scheme}

\begin{figure}
\hspace{-15.0pt}
\includegraphics[width=9cm,height=7.2cm]{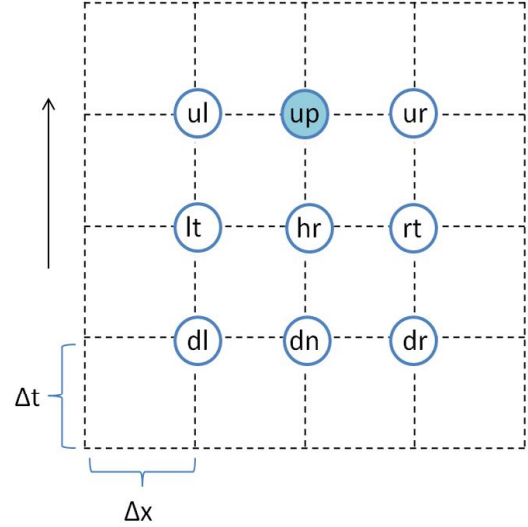}
\caption{Numerical evolution scheme.}
\label{fig:grid_scheme}
\end{figure}

The leapfrog integration scheme is implemented in this paper, which is second-order accurate and nondissipative. With the demonstration of Fig.~\ref{fig:grid_scheme} and using the variable $\phi$ as an example, our discretization scheme is expressed below:
\be
\begin{aligned}
& \frac{d\phi}{dt}=\frac{\phi_{\text{up}}-\phi_{\text{dn}}}{2\Delta t},
& \frac{d\phi}{dx}=\frac{\phi_{\text{rt}}-\phi_{\text{lt}}}{2\Delta x},\\
& \frac{d^{2}\phi}{dt^{2}}=\frac{\phi_{\text{up}}-2\phi_{\text{hr}}+\phi_{\text{dn}}}{(\Delta t)^{2}},
& \frac{d^{2}\phi}{dx^{2}}=\frac{\phi_{\text{lt}}-2\phi_{\text{hr}}+\phi_{\text{rt}}}{(\Delta x)^{2}},\\
& \frac{d^{2}\phi}{dxdt}=\frac{\phi_{\text{ur}}-\phi_{\text{ul}}-\phi_{\text{dr}}+\phi_{\text{dl}}}{4\Delta x\cdot \Delta t},
& \frac{d\phi}{du}=\frac{d\phi}{dt}-\frac{d\phi}{dx}.\\
\end{aligned} \nonumber
\ee
In this paper, we let the temporal and the spatial grid spacings be equal, $\Delta t=\Delta x$.

The equations of motion for $\eta\equiv r^2$ (\ref{equation_r_2}), for $\phi$ (\ref{equation_phi}), and for $\psi$ (\ref{equation_psi}) are coupled. Newton's iteration method can be employed to solve this problem~\cite{Pretorius}. With the illustration of Fig.~\ref{fig:grid_scheme}, the initial conditions provide the data at the levels of \lq\lq down\rq\rq~and \lq\lq here,\rq\rq~and we need to obtain the data on the level of \lq\lq up\rq\rq.~We take the values at the level of \lq\lq here\rq\rq~to be the initial guess for the level of \lq\lq up\rq\rq.~Then, we update the values at the level of \lq\lq up\rq\rq~using the following iteration (taking $\phi$ as an example):
\be \phi^{\text{new}}_{\text{up}}=\phi_{\text{up}}-\frac{G(\phi_{\text{up}})}{J(\phi_{\text{up}})},\nonumber\ee
where ${G}(\phi_{\text{up}})$ is the residual of the differential equation for the function $\phi_{\text{up}}$, and $J(\phi_{\text{up}})$ is the Jacobian defined by
\be J(\phi_{\text{up}})=\frac{\partial G(\phi_{\text{up}})}{\partial \phi_{\text{up}}}.\nonumber\ee
We do the iterations for all of the coupled equations one by one, and run the iteration loops until the desired accuracies are achieved.

\subsection{\texorpdfstring{Locating apparent horizon and examining dynamics near the singularity with mesh refinement}{Horizon and singularity}\label{sec:set_up_locate_AH}}

Horizons are important characteristics of black holes. For simplicity, we locate the apparent horizon of a black hole formed in the collapse, where the expansion of the outgoing null geodesics orthogonal to the apparent horizon is zero~\cite{Baumgarte}. This implies that, in double-null coordinates~\cite{Csizmadia}, at the apparent horizon
\be \tilde g^{\mu\nu}r_{,\mu}r_{,\nu}=e^{2\sigma}(-r_{,t}^2+r_{,x}^2)=1-\frac{2m}{r}=0. \label{locate_AH}\ee
With this property, one can look for the apparent horizon. $M(\equiv m/G)$ in Eq.~(\ref{locate_AH}) is the mass of the black hole.

\begin{figure}
\hspace{-40pt}
\epsfig{file=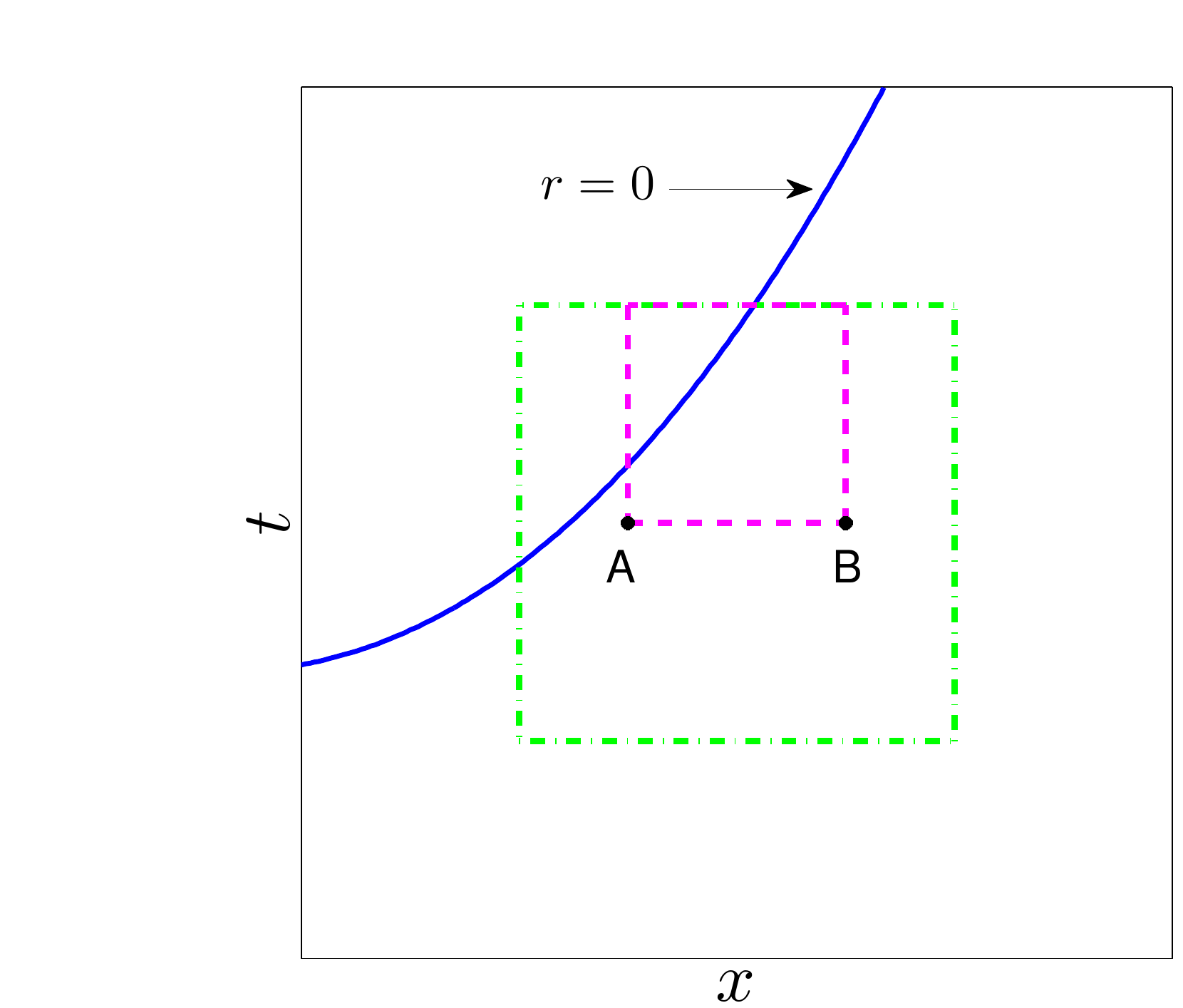, width=8.0cm, height=6.7cm}
\caption{(color online) Description of fixed mesh refinement. Firstly, with numerical results obtained using coarse grid points, we locate the singularity curve $r=0$ roughly, as shown by the solid (blue) line, and choose a region to examine, e.g., the region enclosed by the dash-dotted (green) square. Then, the grid points in this region are interpolated with the original grid spacing being halved. We take two new neighboring slices as initial data for the next simulation. Specifically, the new initial data are located near the line segment $AB$. The interpolate-and-run loop is repeated until the desired accuracies are achieved.}
\label{fig:diagram_FMR}
\end{figure}

Gravity near the singularity is super strong. In order to study the dynamics and examine the BKL conjecture in this region, high-resolution simulations are needed. To achieve this, one may choose to slow down the evolution near the singularity by multiplying the $(t,t)$ metric component with an appropriate lapse function~\cite{Csizmadia}. However, in this paper, we employ an alternative approach: fixed mesh refinement, which is similar to the one used in Ref.~\cite{Garfinkel_3}. This method is very convenient to implement and works very well. Firstly, with numerical results obtained using coarse grid points, we roughly locate the singularity curve $r=0$, as shown by the solid (blue) line in Fig.~\ref{fig:diagram_FMR}, and choose a region to examine, e.g., the region enclosed by the dash-dotted (green) square. Then the grid points in this region are interpolated with the original grid spacing being halved. We take two neighboring slices, with narrower spatial range, of the newly interpolated results at the midway as new initial data. Specifically, the new initial data are located near the line segment $AB$ in Fig.~\ref{fig:diagram_FMR}. We then run the simulations with these new initial data. The interpolate-and-run loop is iterated until the desired accuracies are obtained.

As discussed in Sec.~\ref{sec:set_up_EoM}, in the first simulation with coarse grid points, the term $(-r_{,tt}+r_{,xx})/r$ in (\ref{equation_sigma}) can create big errors near the center $x=r=0$. To avoid this problem, we use the constraint equation (\ref{constraint_eq_uu}) instead. However, at the mesh refinement stage, in the region that we are investigating, the values of $r$ at the two boundaries are usually as regular as those at other neighboring grid points. We need to study the behaviors of all the terms in Eq.~(\ref{equation_sigma}) with high accuracy. Therefore, at the mesh refinement stage, we switch back to Eq.~(\ref{equation_sigma}). The values of the integration variables on the two boundaries are obtained via extrapolation.

\subsection{Numerical tests}
\begin{figure}
  \epsfig{file=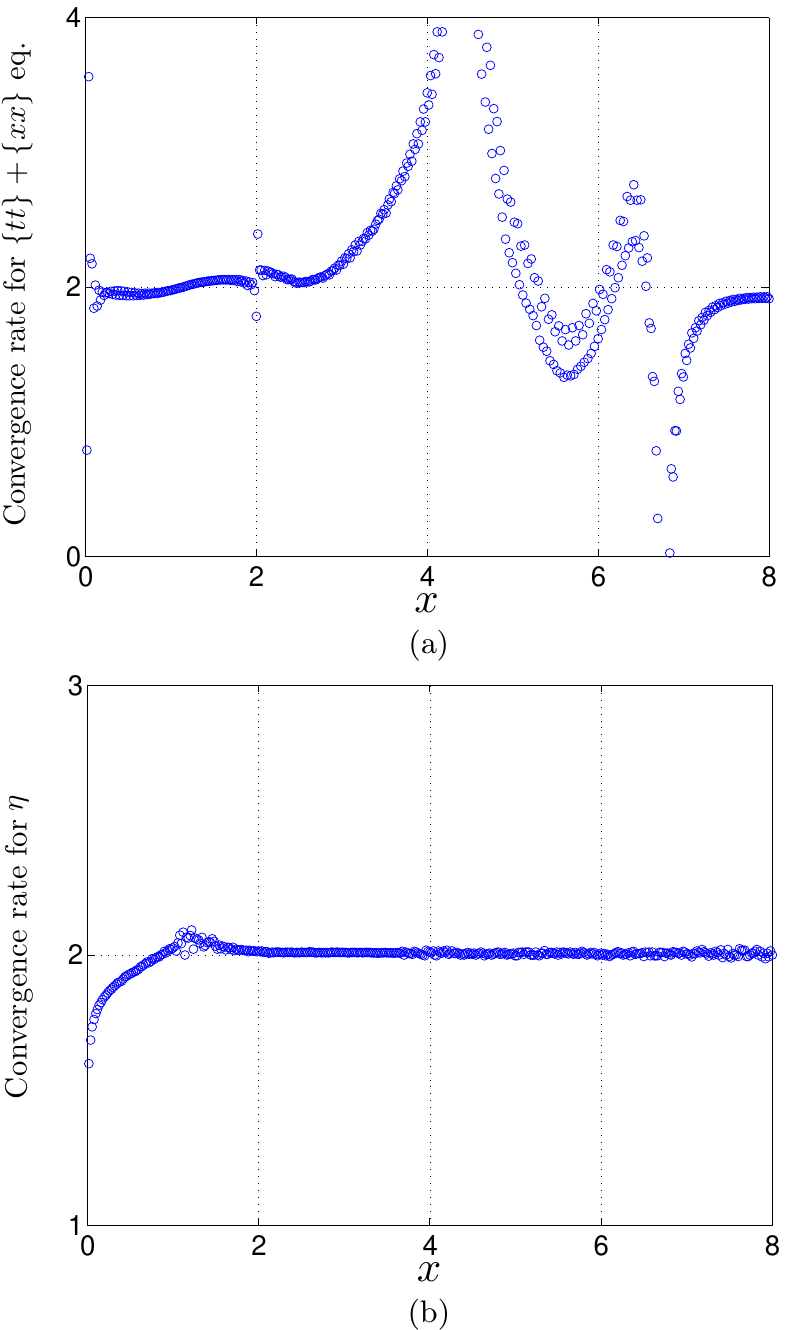, width=8cm,height=13.2cm}
  \caption{Numerical tests when the coordinate time $t$ is equal to $3.5$. (a) Convergence, described by Eq.~(\ref{accuracy_test}), for $\{tt\}+\{xx\}$ constraint equation (\ref{constraint_eq_xx_tt}). In the numerical simulations, this constraint equation is about second-order convergent. (b) Convergence rate, expressed by Eq.~(\ref{convergence_test}), for $\eta\equiv r^2$. The simulation results for $\eta$ are second-order convergent.}
  \label{fig:numerical_tests}
\end{figure}

The accuracies of the discretized equations of motion used in the simulations are checked. In the simulations, the range for the spatial coordinate is $x\in[0 \mbox{ } 22]$, and the grid spacing $\Delta x$ of the coarsest grid is set to $0.01$. The constraint equations (\ref{constraint_eq_xt}) and (\ref{constraint_eq_xx_tt}) are also examined. The convergence rate of a discretized equation can be obtained from the ratio between residuals with two different step sizes,
\be n=\log_{2}\left[\frac{\mathcal{O}(h^{n})}{\mathcal{O}\left(\left(\frac{h}{2}\right)^{n}\right)}\right].\label{accuracy_test}\ee
Our numerical results show that both of the constraint equations are about second-order convergent. As a representative, in Fig.~\ref{fig:numerical_tests}(a), we plot the results for the $\{tt\}+\{xx\}$ constraint equation (\ref{constraint_eq_xx_tt}) when the coordinate time is equal to $3.5$.

Convergence tests via simulations with different grid sizes are also implemented~\cite{Sorkin,Golod}. If the numerical solution converges, the relation between the numerical solution and the real one can be expressed by
\be F_{\mbox{real}}=F^{h}+\mathcal{O}(h^{n}),\nonumber\ee
where $n$ is the convergence order, and $F^{h}$ is the numerical solution with step size $h$. Then, for step sizes equal to $h/2$ and $h/4$, we have
\be F_{\mbox{real}}=F^{\frac{h}{2}}+\mathcal{O}\left[\left(\frac{h}{2}\right)^{n}\right],\nonumber\ee
\be F_{\mbox{real}}=F^{\frac{h}{4}}+\mathcal{O}\left[\left(\frac{h}{4}\right)^{n}\right].\nonumber\ee
Defining $c_1\equiv F^{h}-F^{h/2}$ and $c_2\equiv F^{h/2}-F^{h/4}$, one can obtain the convergence rate
\be n=\log_{2}\left(\frac{c_1}{c_2}\right). \label{convergence_test}\ee
The convergence tests for $\eta\equiv r^{2}$, $g$, $\phi$, and $\psi$ are investigated, and they are all second-order convergent. As a representative, in Fig.~\ref{fig:numerical_tests}(b), the results for $\eta$ are plotted when the coordinate time is equal to $3.5$.

\section{Results\label{sec:results}}
A black hole formation from the scalar collapse in $f(R)$ gravity is obtained. During the collapse, the scalar degree of freedom $f'$ is decoupled from the source scalar field $\psi$ and becomes light. Consequently, gravity transits from general relativity to $f(R)$ gravity. Near the singularity, the contributions of various terms in the equations of motion for the metric components and scalar fields are studied. The asymptotic solutions for the metric components and the scalar field $\phi$ near the singularity are obtained. They are the Kasner solution. These results support the BKL conjecture well.

\begin{figure}[htbp]
  \epsfig{file=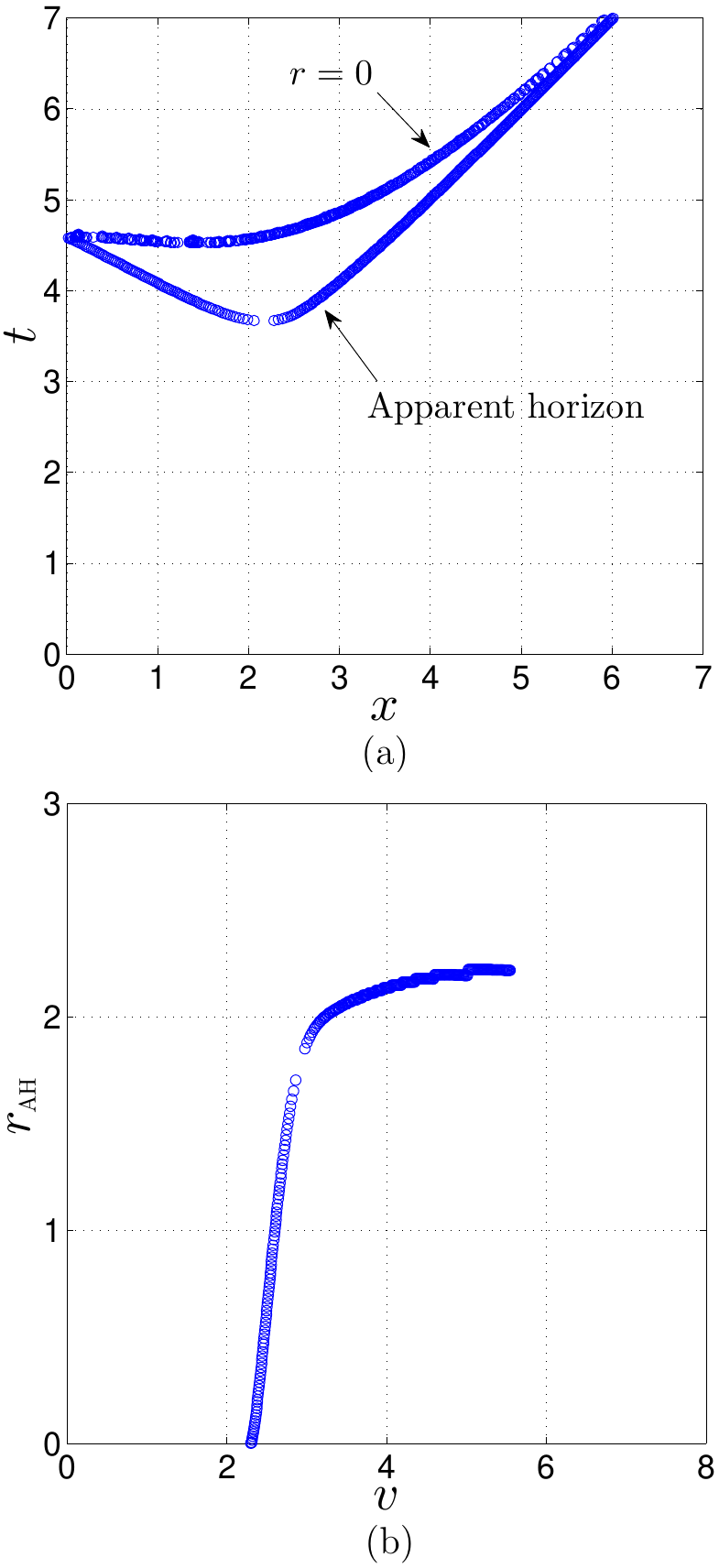, width=6.6cm,height=14.18cm}
  \caption{Apparent horizon of the black hole formed from spherical collapse for the Hu-Sawicki model described by Eq.~(\ref{f_R_Hu_Sawicki}), with $D=1.2$ and $R_{0}=5\times10^{-6}$. $v=(x+t)/2$.}
  \label{fig:apparent_horizon}
\end{figure}

\subsection{Black hole formation}
Before the collapse, near the scalar sphere, $f'$ stays at the right side of the potential (shown in Fig.~\ref{fig:potential_Hu_Sawicki}) due to the balance between $U'(f')$ and the force from the physical scalar field $\psi$. During the collapse, the force from $\psi$ decreases and then changes the direction at a later stage. Correspondingly, $f'$ rolls down the potential and then crosses the minimum of the potential, as depicted in Fig.~\ref{fig:potential_Hu_Sawicki}. If the energy carried by the scalar field $\psi$ is small enough, the field $f'$ will oscillate and eventually stop at the minimum of the potential, and the field $\psi$ disperses. The resulting spacetime is a de Sitter spacetime. However, if the scalar field $\psi$ carries enough energy, a black hole will form, and the Weyl tensor and Weyl scalar will become singular as $r$ goes to zero, which is confirmed in Sec.~\ref{sec:view_JF} and Fig.~\ref{fig:invariants}. This implies that $r=0$ is the true singularity inside a black hole. Moreover, with Eq.~(\ref{locate_AH}), the apparent horizon is found and plotted in Fig.~\ref{fig:apparent_horizon}. Therefore, a black hole is formed.

\begin{figure*}[t!]
  \epsfig{file=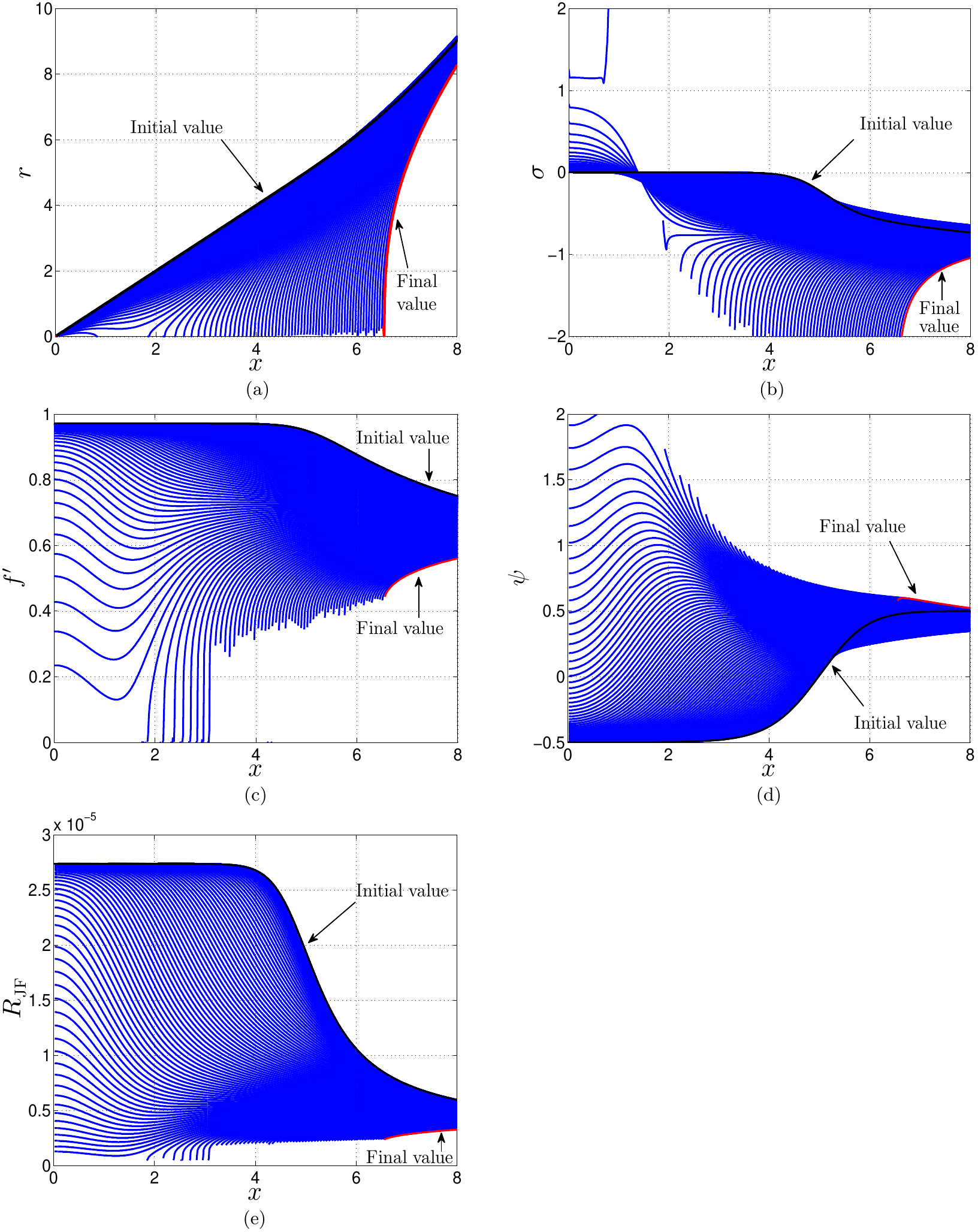, width=17cm,height=20.6cm}
  \caption{Evolutions of the metric components and scalar fields on consecutive time slices. As the singularity curve, $r=0$, is approached, $f'$ goes to zero, and the physical scalar field $\psi$ becomes singular. Near the boundary of the scalar sphere,
  the Ricci scalar in the Jordan frame, $R_{\mbox{\scriptsize JF}}$, moves from a large value at the initial state to a very small value as one moves to the singularity [also refer to Fig.~\ref{fig:invariants}(b)]. Then, gravity transits from general relativity to $f(R)$ gravity.
  }
  \label{fig:Evolution_metric_fields}
\end{figure*}

\begin{figure*}[t!]
  \epsfig{file=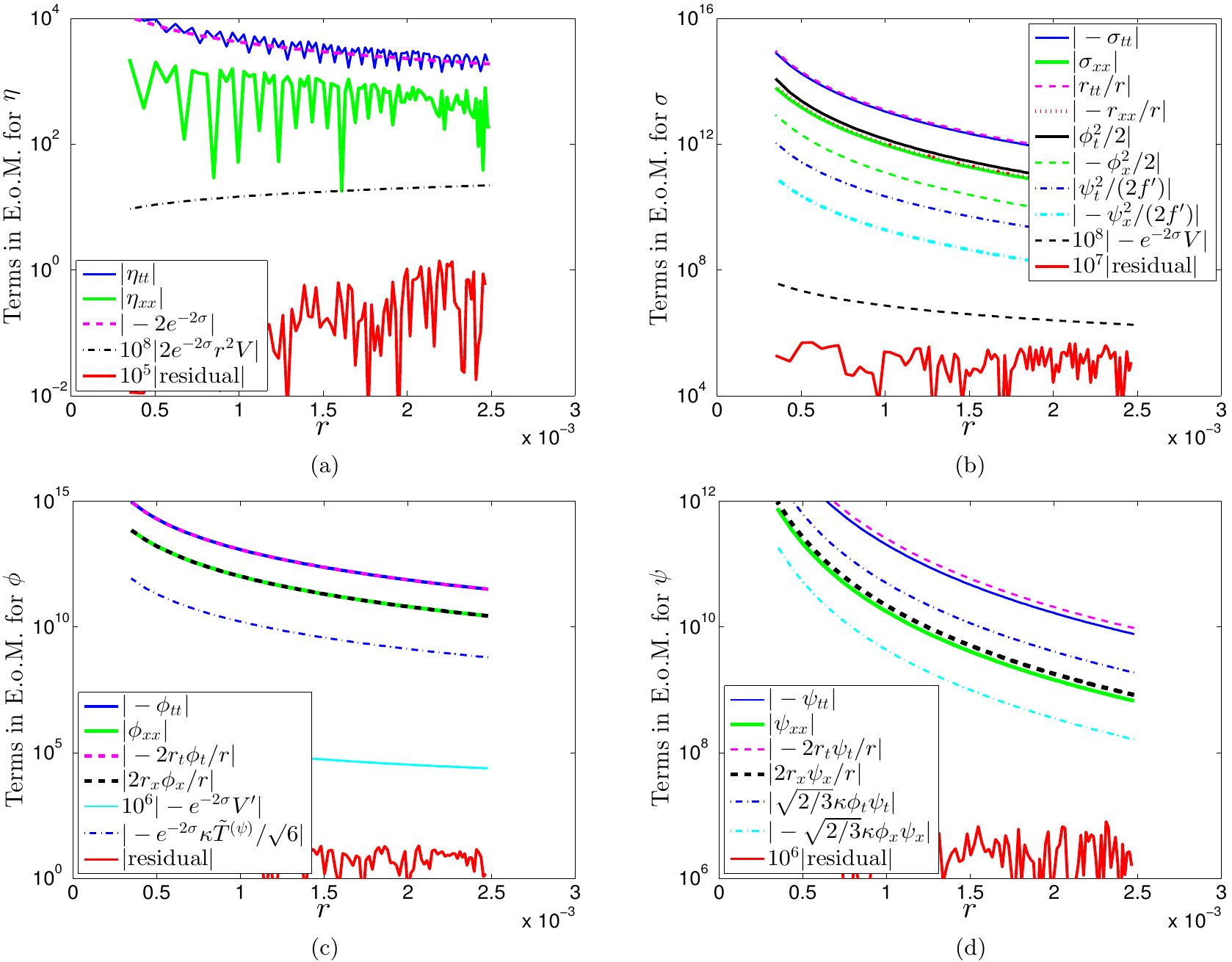, width=17.2cm,height=13.5cm}
  \caption{(color online) Numerical solutions near the singularity obtained via mesh refinement. The range for the spatial coordinate is $x\in[0 \mbox{ } 22]$, and the solution shown in this figure is for $(x=2.5,t=t)$. In the vicinity of the singularity, in the equations of motion, the metric component terms are the most important, and the potential terms are the least important. The scalar fields are intermediate. The scalar field $\phi$ dominates over the physical field $\psi$. The ratios of $\sigma_{,tt}/\sigma_{,xx}$, $\phi_{,t}^2/\phi_{,x}^2$, and $(r_{,t}\phi_{,t})/(r_{,x}\phi_{,x})$ are all around $11.5$. As discussed in Sec.~\ref{sec:results_dynamics}, this is related to the slope of the singularity curve. This implies that the slope of the singularity curve at $x=2.5$ is about $\sqrt{1/11.5}\approx 0.29$. Consequently, neglecting minor terms, we can approximately rewrite the original equations of motion for $\sigma$, $\phi$, and $\psi$ only in terms of temporal derivatives.
  (a) The equation of motion for $\eta$ (\ref{equation_r_2}) becomes $-\eta_{,tt}+\eta_{,xx}\approx 2e^{-2\sigma}$.
  (b) The equation of motion for $\sigma$ (\ref{equation_sigma}) becomes $-\sigma_{,tt}+r_{,tt}/r+\phi_{,t}^2/2\approx 0$.
  (c) The equation of motion for $\phi$ (\ref{equation_phi}) becomes $\phi_{,tt}+2r_{,t}\phi_{,t}/{r}\approx 0$.
  (d) The equation of motion for $\psi$ (\ref{equation_psi}) becomes $\psi_{,tt}+2r_{,t}\psi_{,t}/{r}\approx \sqrt{2/3}\kappa\phi_{,t}\psi_{,t}$. Note that $\phi_{,t}$ is negative. Therefore, the term $\sqrt{2/3}\kappa\phi_{,t}\psi_{,t}$ tries to stop the evolution of $\psi$. This is a dark energy effect.
  }
  \label{fig:EoM_sigulairty}
\end{figure*}

\begin{figure}[h]
\hspace{-10pt}
\epsfig{file=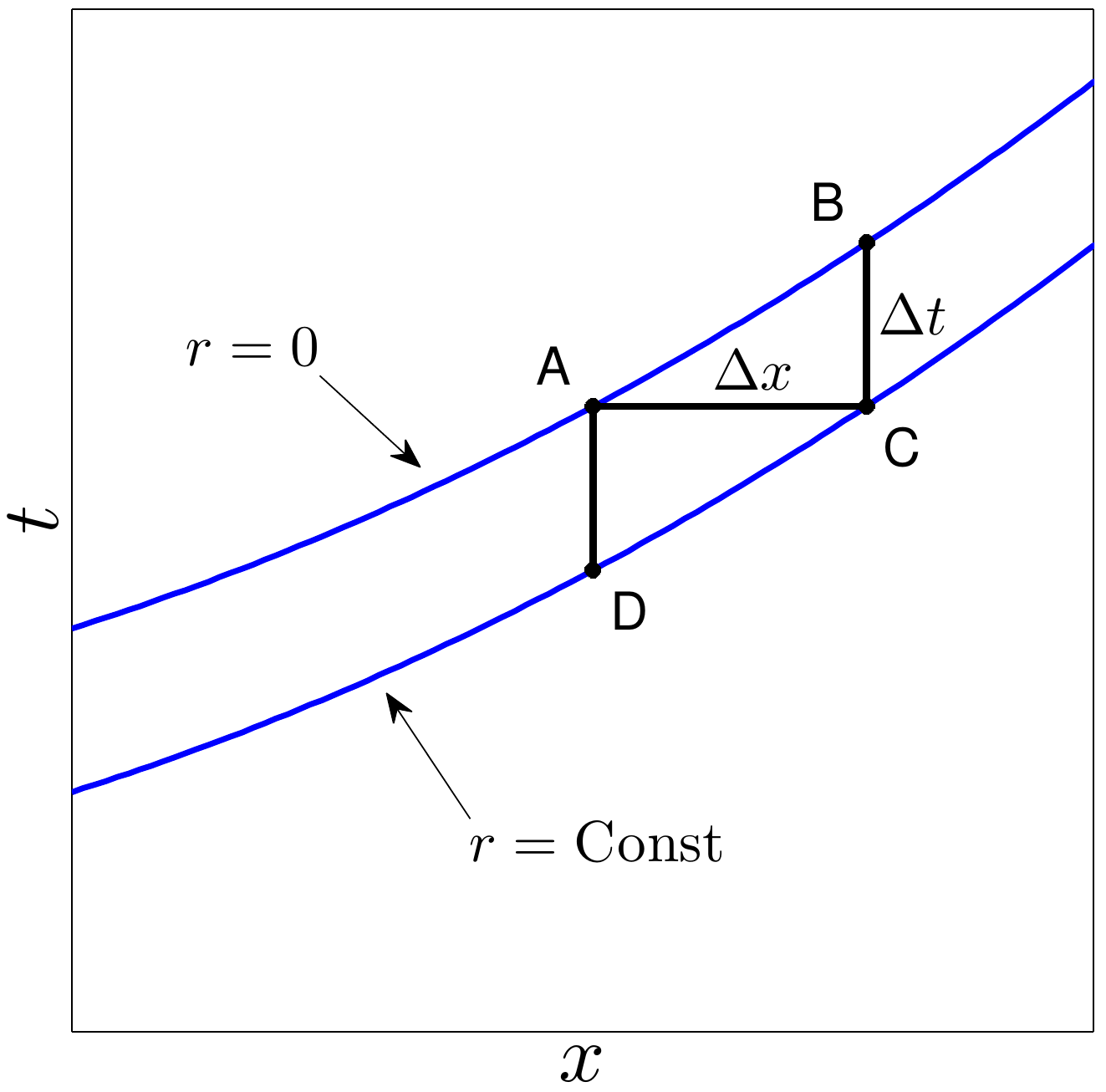, width=6.3cm, height=6.2cm}
\caption{Spatial derivative vs temporal derivative near the singularity. Point $A$ and point $B$ are on one same hypersurface $r=\mbox{Const}$, while point $C$ is on another one. At point $C$, in first-order accuracy, $r_{,x}\approx(r_{C}-r_{A})/\Delta x$ and $r_{,t}\approx(r_{B}-r_{C})/\Delta t$. Since $r_{A}=r_{B}$ and the slope of the singularity curve, $dt/dx$, is less than $1$, there is $|r_{,x}/r_{,t}|\approx|\Delta t/\Delta x|<1$.}
\label{fig:spatial_vs_temporal}
\end{figure}

\subsection{Dynamics during collapse\label{sec:results_dynamics}}
\begin{figure*}[t!]
  \epsfig{file=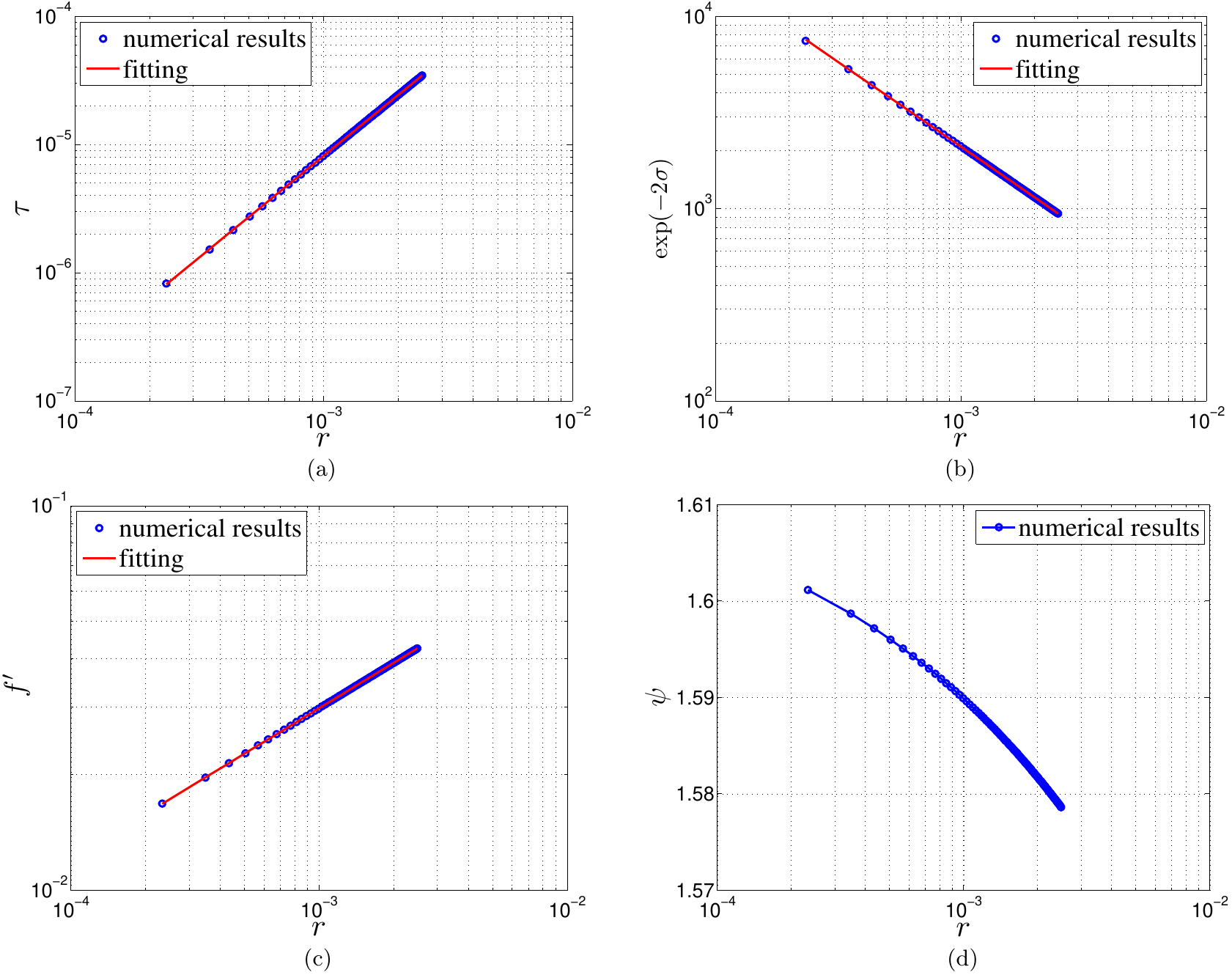, width=17cm,height=13.6cm}
  \caption{Evolutions of the metric components and scalar fields near the singularity obtained via mesh refinement. The range for the spatial coordinate is $x\in[0 \mbox{ } 22]$, and the results shown in this figure are for $(x=2.5,t=t)$. We fit the results near the singularity as follows.
(a) $\log\tau=a+b\log r$, $a=-0.774\pm0.004$, $b=1.5843\pm0.0006$.
(b) $\log e^{-2\sigma}=-2\sigma=a+b\log r$, $a=1.575\pm0.003$, $b=-0.8797\pm0.0004$. $\log e^{-2\sigma}(=-2\sigma)$ has an ideal linear relation with $\log r$, which supports our statements on Eqs.~(\ref{r_asymptotic}) and (\ref{sigma_asymptotic}).
(c) $\log f'=\sqrt{2/3}\kappa \phi=a+b\log r$, $a=-0.8021\pm0.0002$, $b=0.39288\pm0.00004$. Near the singularity, the dynamics of $f'$ or $\phi$ is mainly determined by gravity. Then, $\log f'(=\sqrt{2/3}\kappa \phi)$ has an ideal linear relation with $\log r$. $f'$ approaches zero as $r$ goes to zero.
(d) $\psi$. Near the singularity, although the evolution of $\psi$ is mainly determined and accelerated by gravity, it is considerably suppressed by $\phi$. Then, $\psi$ does not have an ideal linear relation with $\log r$.}
  \label{fig:Kasner_metric_fields}
\end{figure*}

The evolutions of $r$, $\sigma$, $f'$, $\psi$, and the Ricci scalar in the Jordan frame, $R_{\mbox{\scriptsize JF}}$, are shown in Fig.~\ref{fig:Evolution_metric_fields}. During the collapse, the major part of the energy of the source scalar field $\psi$ is transported to the center. Consequently, the field $f'$ is decoupled from the source field and becomes light. At the same time, as shown in Fig.~\ref{fig:Evolution_metric_fields}(e), the Ricci scalar in the Jordan frame decreases, and the modification term in the function $f(R)$ becomes important. In this process, gravity transits from general relativity to $f(R)$ gravity. Compared to gravity from the singularity, the left side of the potential $U(f')$ is not steep enough to stop $f'$ from running to the left. Consequently, the field $f'$ rolls down from its initial value, which is close to $1$, crosses the de Sitter point, and asymptotes to but does not cross zero near the singularity, as shown in Figs.~\ref{fig:Evolution_metric_fields}(c) and~\ref{fig:Kasner_metric_fields}(c). Simultaneously, as shown in Eq.~(\ref{T_psi}), the factor $1/f'$ accelerates the transformed energy-momentum of the source field $\psi$ in the Einstein frame to blow up. In other words, one may say that the effective gravitational coupling constant becomes singular at this point. The observations that $f'$ approaches zero are consistent with the results of collapse in Brans-Dicke theory obtained in Ref.~\cite{Hwang_2010}. One may take $f(R)$ theory as the $\omega=0$ case of Brans-Dicke theory, where $\omega$ is the Brans-Dicke coupling constant. On the other hand, the potential in $f(R)$ theory has a more complicated form than in Brans-Dicke theory. In the latter case, the potential is usually set to zero.

We examine the evolutions in the vicinity of the singularity using fixed mesh refinement as discussed in Sec.~\ref{sec:set_up_locate_AH}. On the sample slice $(x=2.5,t=t)$ that we choose to study, the interpolate-and-run loop is iterated $20$ times. As a result, the grid spacing $\Delta x=\Delta t$ is reduced from $10^{-2}$ to $10^{-8}$. The smallest value for the radius $r$ we can reach is reduced from $10^{-2}$ to $10^{-4}$ (see Figs.~\ref{fig:EoM_sigulairty} and \ref{fig:Kasner_metric_fields}). Note that the radius of the apparent horizon of the formed black hole is about $2.2$ [see Fig.~\ref{fig:apparent_horizon}(b)]. The results obtained via mesh refinement support the BKL conjecture well, as discussed below.

One statement of the conjecture is that, in the vicinity of the singularity, gravity dominates over matter fields. This is verified by the results plotted in Fig.~\ref{fig:EoM_sigulairty}. The results show that the metric terms are the most important ones, while the potential term and the effective force term based on the first-order derivative of the potential with respect to the scalar field $\phi$ are the least important. The terms related to the scalar fields are intermediate. The field $\phi$, transformed from the scalar degree of freedom $f'$, dominates the competition between $\phi$ and the physical source field $\psi$ [see Figs.~\ref{fig:EoM_sigulairty}(b)-(d)]. As discussed in the next paragraph, $|\psi_{,t}|$ is no less than $|\psi_{,x}|$. Then, in the equation of motion for $\phi$ (\ref{equation_phi}), the contribution from $\psi$,
$e^{-2\sigma}\kappa \tilde T^{(\psi)}/\sqrt{6}\left[=(\psi_{,t}^2-\psi_{,x}^2)/(\sqrt{6}f')\right]$, is positive. Namely, $\psi$ accelerates the evolution of $\phi$. On the other hand, this contribution is tiny compared to gravity [see Fig.~\ref{fig:EoM_sigulairty}(c)]. The effective force term from the potential is even less than the contribution from $\psi$. This implies that, in the vicinity of the singularity, $\phi$ or $f'$ becomes almost massless. Regarding the equation of motion for $\psi$ (\ref{equation_psi}), the contribution from $\phi$, $\sqrt{2/3}\kappa \phi_{,t}\psi_{,t}$, is relatively important [see Fig.~\ref{fig:EoM_sigulairty}(d)]. In fact, $\phi_{,t}$ is negative. Therefore, the term $\sqrt{2/3}\kappa \phi_{,t}\psi_{,t}$ functions as a friction force for $\psi$. This is a dark energy effect. This effect can also be observed via comparison of Figs.~\ref{fig:Kasner_metric_fields}(c) and (d). Because the dynamics of $\phi$ is mainly determined by gravity, $\phi[\equiv (\sqrt{3/2}\log f')/\kappa]$ has a good linear relation with $\log r$ [also refer to Eqs.~(\ref{r_asymptotic}) and (\ref{phi_asymptotic})]. However, because of the suppression from $\phi$, the field $\psi$ does not have such a linear relation with $\log r$.

The second statement of the BKL conjecture is that, near the singularity, the terms containing temporal derivatives are dominant over those containing spatial derivatives. However, in double-null and Kruskal coordinates, temporal derivatives and spatial derivatives are connected by the slope of the singularity curve. We first take the variable $r$ as an example.
\break
\break
\break
As illustrated in Fig.~\ref{fig:spatial_vs_temporal}, point $A$ and point $B$ are on one same hypersurface $r=\mbox{Const}$, while point $C$ is on another one. At point $C$, in first-order accuracy, $r_{,x}\approx(r_{C}-r_{A})/\Delta x$ and $r_{,t}\approx(r_{B}-r_{C})/\Delta t$. Since $r_{A}=r_{B}$ and the slope of the singularity curve, $dt/dx$, is no greater than $1$ [see Fig.~\ref{fig:apparent_horizon}(a)], there is
\be \Big|\frac{r_{,x}}{r_{,t}}\Big|\approx\Big|\frac{\Delta t}{\Delta x}\Big|<1.\label{ratio_spatial_temporal}\ee
Namely, in the vicinity of the singularity curve, the ratio between spatial and corresponding temporal derivatives is defined by the slope of this singularity curve. (Similar results for a Schwarzschild black hole in Kruskal coordinates can be obtained analytically. Details are given in Appendix~\ref{sec:appendix_B}.) This can also be interpreted in the following way. In double-null and Kruskal coordinates, the time vector is not normal to the hypersurface of $r=\mbox{Const}$. Then, the derivatives in the radial direction have nonzero projections on both hypersurfaces of $x=\mbox{Const}$ and $t=\mbox{Const}$. With Eq.~(\ref{ratio_spatial_temporal}), along a certain slice $(x=\mbox{Const},t=t)$, near the singularity, the ratio between spatial and corresponding temporal derivatives is almost constant. This is also valid for other quantities, e.g., $\sigma$, $\phi$, and $\psi$. This can be explained as follows. We take the scalar field $\phi$ as an example. With the illustration of Fig.~\ref{fig:spatial_vs_temporal}, as this scalar field moves toward the center $r=0$ along the radial direction, two neighboring points on this scalar wave $\phi$ should take close values when they cross points $C$ and $D$, respectively, on one same hypersurface $r=\mbox{Const}$ at two consecutive moments, because these two points on the scalar wave are neighbors and the
\lq\lq distances\rq\rq~$AD$ and $BC$ are more important for their values than the difference between these two neighboring points. In other words, in the vicinity of the singularity curve, gravity is more important than the difference between neighboring points on the scalar wave. These arguments are also supported by numerical results. Near the singularity, the evolution of $\phi$ is described by Eq.~(\ref{phi_asymptotic}): $\phi\approx C\log \xi$, where $\xi$ is the distance between two hypersurfaces of $r=\mbox{Const}$ and $r=0$. In Fig.~\ref{fig:spatial_vs_temporal}, $\xi$ means $AD$ and $BC$. As shown in Fig.~\ref{fig:variation_singularity_curve}(f), the parameter $C$ changes slowly along the singularity curve, compared to the dramatic running of $\log\xi$ near the singularity. We also checked variations of $C$ as $\xi$ takes different scales on one same slice $(x=\mbox{Const},t=t)$. The results show that $C$ also changes very slowly.

On the slice $(x=2.5,t=t)$ that we study, near the singularity, the ratios between second-order temporal derivatives (or the squared/multiplication of first-order time derivatives) and the corresponding spatial derivatives present in Eqs.~(\ref{equation_r_2})-(\ref{equation_psi}), e.g., $\sigma_{,tt}/\sigma_{,xx}$, $\phi_{,t}^2/\phi_{,x}^2$, and $(r_{,t}\phi_{,t})/(r_{,x}\phi_{,x})$, are all around $11.5$. As argued in the above paragraph, this implies that the slope of the singularity curve at $x=2.5$ is about $\sqrt{1/11.5}\approx 0.29$. In addition, as illustrated in Fig.~\ref{fig:EoM_sigulairty}, the term $2e^{-2\sigma}r^{2}V$ in Eq.~(\ref{equation_r_2})
and the terms $e^{-2\sigma}[V'(\phi)+\kappa \tilde T^{(\psi)}/\sqrt{6}]$ in Eq.~(\ref{equation_phi}) are negligible. Consequently, we can approximately rewrite the original equation of motion for $\eta$ (\ref{equation_r_2}) in the format of (\ref{equation_eta_Kasner}), and rewrite the original equations of motion for $\sigma$ (\ref{equation_sigma}), $\phi$ (\ref{equation_phi}), and $\psi$ (\ref{equation_psi}) only in terms of temporal derivatives as follows:
\be -\eta_{,tt}+\eta_{,xx} \approx 2e^{-2\sigma}, \label{equation_eta_Kasner} \ee
\be
-\sigma_{,tt} + \frac{r_{,tt}}{r} + \frac{1}{2}\phi_{,t}^2 \approx 0,
\label{equation_sigma_Kasner}
\ee
\be
\phi_{,tt}+\frac{2r_{,t}\phi_{,t}}{r}\approx 0 \Longleftrightarrow \phi_{,t}\approx \text{Const}\cdot r^{-2}+\text{Const},
\label{equation_phi_Kasner}
\ee
\be
\psi_{,tt}+\frac{2r_{,t}\psi_{,t}}{r} \approx \sqrt{\frac{2}{3}}~\kappa \phi_{,t}\psi_{,t}.
\label{equation_psi_Kasner}
\ee
Note that $|\eta_{,xx}|$ is no greater than $|\eta_{,tt}|$. As the singularity is approached, $r_{,t}$ and $\phi_{,t}$ are both negative. (Refer to the above arguments at the beginning of this section.) Then, Eq.~(\ref{equation_phi_Kasner}) implies that $\phi_{,tt}<0$. Therefore, $\phi$ will be accelerated to $-\infty$. Correspondingly, $f'$ approaches zero. Similar arguments can be applied to other equations above. Then, the dynamical system approaches an attractor $(r\rightarrow0,\sigma=-\infty,f'\rightarrow 0,\psi=+\infty)$. Next, we will explore the asymptotic solutions based on Eqs.~(\ref{equation_eta_Kasner})-(\ref{equation_phi_Kasner}).

\subsection{\texorpdfstring{Kasner solution for Schwarzschild black hole}{Kasner solution for Schwarzschild BH}}
The third statement of the BKL conjecture is that the dynamics near the singularity is expressed by the universal Kasner solution~\cite{Kasner}.
The four-dimensional homogeneous but anisotropic Kasner solution with a massless scalar field $\zeta$ minimally coupled to gravity can be described as follows~\cite{Kamenshchik,Nariai,Belinskii_2}:
\be
\begin{array}{l l}
  ds^2=-d\tau^2+\sum\limits_{i=1}^3 \tau^{2p_i}dx_{i}^{2},\\
  \\
  p_1+p_2+p_3=1,\\
 \\
  p^{2}_{1}+p^{2}_{2}+p^{2}_{3}=1-q^2,\\
  \\
  \zeta=q\log \tau,
\end{array}
\label{Kasner_solution}
\ee
where the parameter $q$ describes the contribution from the field $\zeta$. The parameter $q^2$ is constrained by Eq.~(\ref{Kasner_solution}) as
\be q^2 \leq \frac{2}{3}.\ee
The Kasner exponents can be expressed in the following parametric form:
\be p_1=\frac{-w}{1+w+w^2},\label{kasner_p1}\ee
\be p_2=\frac{1+w}{1+w+w^2}\left\{w-\frac{w-1}{2}\left[1-(1-\alpha^2)^{\frac{1}{2}}\right]\right\},
\label{kasner_p2}\ee
\be p_3=\frac{1+w}{1+w+w^2}\left\{1+\frac{w-1}{2}\left[1-(1-\alpha^2)^{\frac{1}{2}}\right]\right\},
\label{kasner_p3}\ee
\be \alpha^2=\frac{2(1+w+w^2)^2q^2}{(w^2-1)^2}.\label{kasner_alpha}\ee
The parameter $\alpha^2$ is no greater than 1. The Kasner exponents are invariant under the transformation of $w\rightarrow 1/w$:
\be
\begin{array}{l l}
  p_1\left(\frac{1}{w}\right)=p_1(w),\\
  \\
  p_2\left(\frac{1}{w}\right)=p_3(w),\\
  \\
  p_3\left(\frac{1}{w}\right)=p_2(w).
\end{array}
\nonumber
\ee
If $q^2>0$, there are combinations of positive Kasner exponents, satisfying Eq.~(\ref{Kasner_solution}). Moreover, all three Kasner exponents take positive values if $q^2\geq 1/2$~\cite{Kamenshchik,Belinskii_2}. As will be demonstrated in the rest of this paper, a Schwarzschild black hole and spherical collapse toward a black hole formation have special types of Kasner solution, in which $p_2$ and $p_3$ are equal.

The behavior of a test scalar field near the singularity in the spacetime of the Oppenheimer-Snyder collapse~\cite{Oppenheimer} was simulated in Ref.~\cite{Garfinkel_2}. The spacetime is asymptotically flat. The results confirmed one statement of the BKL conjecture: the temporal derivative terms are dominant over the spatial ones. In the scalar collapse in $f(R)$ gravity that we study in this paper, two scalar fields are present. One of them, $\psi$, is massless, and the other one, $\phi$, is very light although it has a mass. Moreover, the spacetime has an asymptotic de Sitter solution.

Due to the close connection between a Schwarzschild black hole and spherical collapse, it is instructive to review the dynamics near the singularity of   a Schwarzschild black hole first. In Schwarzschild coordinates, the Schwarzschild metric can be expressed as
\be ds^2=-\left(1-\frac{2m}{r}\right)dt^2+\frac{1}{1-\frac{2m}{r}}dr^2+r^2d\Omega^2,\nonumber\ee
which, near the singularity, is reduced to
\be ds^2\approx-\frac{r}{2m}dr^2+\frac{2m}{r}dt^2+r^2d\Omega^2.\label{Schwarzschild_singularity}\ee
Inside the horizon, $r$ is timelike, and $t$ is spacelike. In this case,
\be \tau\approx\int_0^r \sqrt{\frac{r}{2m}}dr=\frac{\sqrt{2}}{3\sqrt{m}}r^{\frac{3}{2}},
  \hphantom{ddd} r\approx\left(\frac{3\sqrt{2m}}{2}\tau\right)^{\frac{2}{3}}.\label{r_Schwarzschild}\ee
Considering Eqs.~(\ref{Kasner_solution}), (\ref{Schwarzschild_singularity}), and (\ref{r_Schwarzschild}), we have
\be p_1=-\frac{1}{3}, \hphantom{ddd} p_2=p_3=\frac{2}{3},\ee
which clearly are Kasner exponents, satisfying Eq.~(\ref{Kasner_solution}), with $q$ being equal to zero.

To be one more step closer to spherical collapse in double-null coordinates, we consider the Schwarzschild metric in Kruskal coordinates, which has the following form:
\be ds^2=\frac{32m^3}{r}e^{-\frac{r}{2m}}(-dt^2+dx^2)+r^2d\Omega^2.\label{Kruskal_coordinate}\ee
The Schwarzschild radius $r$ is given by
\be t^2-x^2=\left(1-\frac{r}{2m}\right)e^{\frac{r}{2m}}.\label{radius_schw_BH}\ee
In the vicinity of the singularity curve, we rewrite $t$ as $t=t_0-\xi$, where $t_0$ is the coordinate time on the singularity curve and $\xi\ll t_0$. With the spatial coordinate $x$ being fixed, a perturbation expansion near the singularity curve directly yields
\be r\approx\left(16m^{2}t_0\xi\right)^{\frac{1}{2}}.\label{r_Kruskal}\ee
Consequently, the proper time is
\be \tau\approx\left(32m^3\right)^{\frac{1}{2}}\int_0^\xi r^{-\frac{1}{2}}d\xi
\approx \frac{8\sqrt{2}m}{3(t_0)^{\frac{1}{4}}}\xi^{\frac{3}{4}}
\approx \frac{\sqrt{2}}{3\sqrt{m}t_0}r^{\frac{3}{2}}.\nonumber \ee
Therefore,
\be r\approx\left(\frac{3\sqrt{2m}t_0}{2}\tau\right)^{\frac{2}{3}}. \ee
Then, we obtain the same set of Kasner exponents as in Schwarzschild coordinates.

\subsection{Kasner solution for spherical collapse}
The reduced equations of motion, (\ref{equation_eta_Kasner})-(\ref{equation_phi_Kasner}), numerical results for spherical collapse in $f(R)$ theory, and the analysis of dynamics near the singularity for a Schwarzschild black hole together show that the variables $r$, $\sigma$, and $\phi$ have the following asymptotic solutions:
\be r\approx A\xi^{\beta},\label{r_asymptotic}\ee
\be \sigma\approx B\log\xi,\label{sigma_asymptotic}\ee
\be \phi\approx C\log\xi,\label{phi_asymptotic}\ee
where $\xi$ is defined in the same way as in the Kruskal case: $\xi=t_0-t$, where $t_0$ is the coordinate time on the singularity curve. Substituting the above three expressions into Eq.~(\ref{equation_sigma_Kasner}) yields a relation between parameters $\beta$, $B$, and $C$:
\be B\approx\beta(1-\beta)-\frac{C^2}{2}.\label{sigma_B}\ee
We then put Eqs.~(\ref{r_asymptotic}), (\ref{sigma_asymptotic}), and (\ref{sigma_B}) into (\ref{equation_eta_Kasner}). Noting that the ratio $\eta_{,tt}/\eta_{,xx}$ has a certain value near a fixed singularity point, and neglecting minor terms, we obtain
\be \log(1-2\beta)\approx\left[2(\beta-1)^2+C^2\right]\log\xi.\nonumber\ee
As $\xi$ approaches zero, the parameter $\beta$ needs to be close to $1/2$, so that the two sides of the equation are balanced. In this case, the above equation implies that
\be \beta\approx \frac{1-\xi^{\frac{1}{2}+C^2}}{2}.\label{beta_kasner}\ee
Therefore, as a function of $\xi$, $r$ in spherical collapse has an exponent close to the one in a Schwarzschild black hole in Kruskal coordinates [see Eq.~(\ref{r_Kruskal})]. Substitution of Eqs.~(\ref{sigma_B}) and (\ref{beta_kasner}) into (\ref{sigma_asymptotic}) leads to the asymptotic solution for $\sigma$,
\be \sigma\approx\left(\frac{1-2C^2}{4}\right)\log\xi.\label{sigma_asymptotic_2}\ee
Then, the proper time is
\be \tau=\int_0^{\xi}e^{-\sigma}d\xi\approx\frac{4}{3+2C^2}\xi^{\frac{3+2C^2}{4}}.\ee
Consequently, one can obtain the expressions for the metric components and scalar field $\phi$ with respect to $\tau$ as follows:
\be r\approx A\xi^{\frac{1}{2}}\approx A\left(\frac{3+2C^2}{4}\tau\right)^{\frac{2}{3+2C^2}}, \label{r_kasner}\ee
\be e^{-\sigma}\approx\xi^{\frac{-1+2C^2}{4}}\approx\left(\frac{3+2C^2}{4}\tau\right)^{\frac{-1+2C^2}{3+2C^2}},
\label{a_sigma_kasner}\ee
\be \phi\approx C\log\xi\approx\frac{4C}{3+2C^2}\log\tau.\label{phi_kasner}\ee

\begin{figure*}[t!]
  \epsfig{file=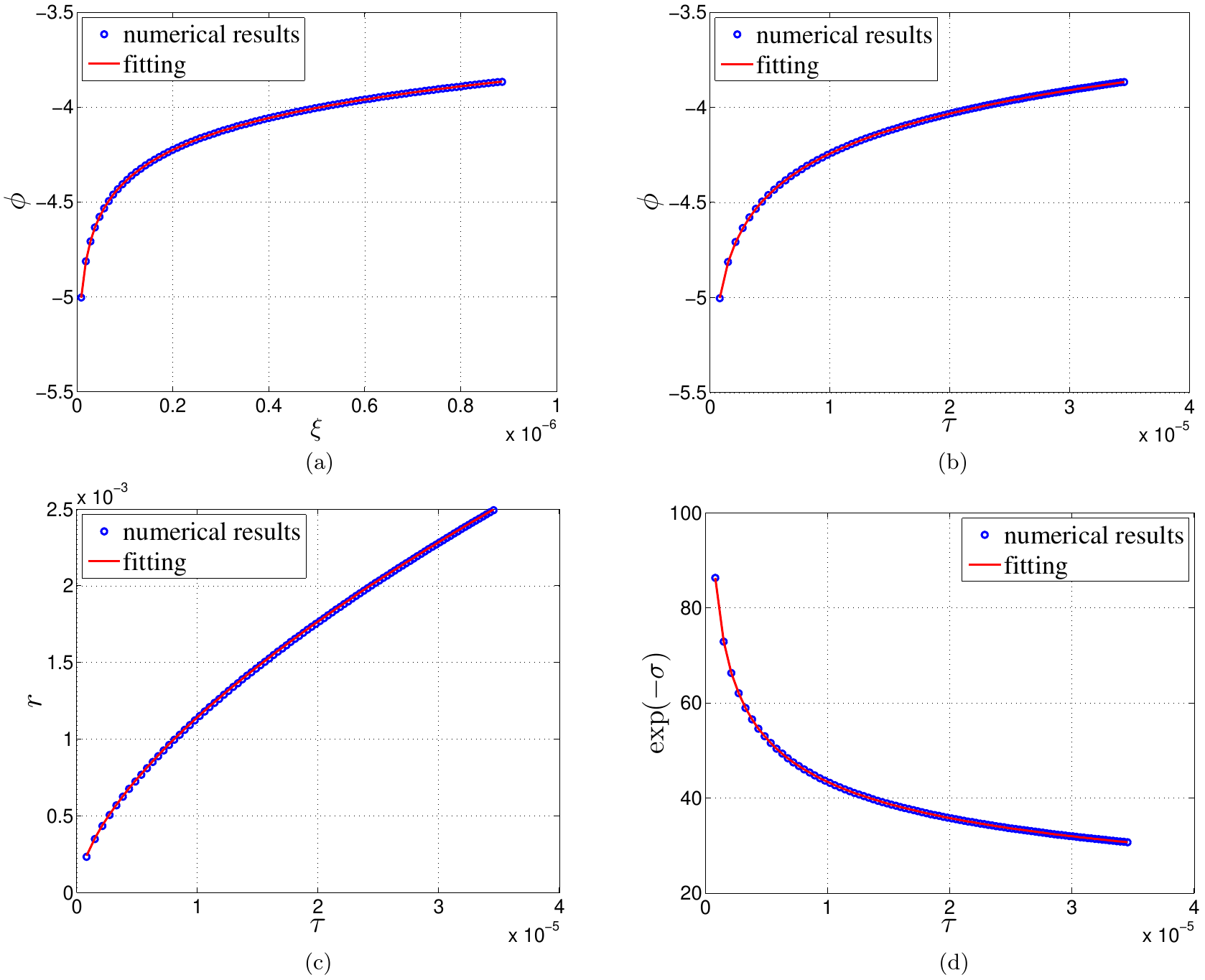, width=17cm,height=13.8cm}
  \caption{Verification of the Kasner solution near the singularity. The results shown in this figure are obtained via mesh refinement. These results are for $(x=2.5,t=t)$. We fit the results near the singularity as follows.
(a) $\phi=a+b\log(\xi+c)$, $a=-0.5118\pm0.0004$, $b=0.24070\pm0.00003$, $c=(-1.735\pm 0.008)\times10^{-9}$.
(b) $\phi=a+b\log(\tau+c)$, $a=-0.747\pm0.002$, $b=0.3038\pm0.0002$, $c=(-0.89\pm2.75)\times10^{-9}$.
(c) $r=a+b(\tau+c)^{d}$, $a=(6.00\pm0.06)\times10^{-5}$, $b=1.895\pm0.002$, $c=(-2.17\pm0.03)\times10^{-7}$, $d=0.6475\pm0.0002$.
(d) $\exp(-\sigma)=a+b(\tau+c)^{d}$, $a=-1.98\pm0.05$, $b=2.14\pm0.01$, $c=(-1.24\pm0.09)\times10^{-8}$, $d=-0.2650\pm0.0003$.}
  \label{fig:Kasner_fit}
\end{figure*}

Comparing Eqs.~(\ref{r_kasner})-(\ref{phi_kasner}) to (\ref{Kasner_solution}), we extract
\be p_1=\frac{-1+2C^2}{3+2C^2}, \hphantom{ddd} p_2=p_3=\frac{2}{3+2C^2},\label{kasner_parameters}\ee
and
\be q=\frac{4C}{3+2C^2}.\label{kasner_q}\ee
It can be verified that these parameters satisfy Eq.~(\ref{Kasner_solution}). It is noticeable that as the parameter $C$ in Eq.~(\ref{phi_kasner}) goes to zero, namely the field $\phi$ disappears, the Kasner exponents take the same values as in the Schwarzschild black hole case. The above analytic expressions are also supported by numerical results. On the slice that we study, the parameter $C$ for $\phi$ is obtained by fitting the numerical results, $C=0.24070\pm0.00003$ [see Fig.~\ref{fig:Kasner_fit}(a)]. Then, with Eqs.~(\ref{kasner_parameters}) and (\ref{kasner_q}), the values for the Kasner exponents and the parameter $q$ are
\be
\begin{array}{l l}
  p_1=-0.28375\pm0.00001,\\
  \\
  p_2=p_3=0.641874\pm0.000006,\\
  \\
  q=0.308998\pm0.000003.
\end{array}
\nonumber
\ee
As shown in Figs.~\ref{fig:Kasner_fit}(b)-(d), the values for these quantities obtained via fitting the numerical results are
\be
\begin{array}{l l}
  p_1=-0.2650\pm0.0003,\\
  \\
  p_2=p_3=0.6475\pm0.0002,\\
  \\
  q=0.3038\pm0.0002.
\end{array}
\nonumber
\ee
The two sets of values are highly compatible. Therefore, we obtain the Kasner solution for spherical scalar collapse in $f(R)$ theory in double-null coordinates in the Einstein frame.

\subsection{Variations of Kasner parameters along the singularity curve}
The Kasner solution described by Eqs.~(\ref{kasner_parameters}) and (\ref{kasner_q}) is a special case of the one expressed by Eqs.~(\ref{kasner_p1})-(\ref{kasner_alpha}). The two sets of expressions are identical under the conditions
\be \alpha^2=1, \hphantom{ddd} |C|=\frac{1}{\sqrt{2}}\left|{\frac{1-w}{1+w}}\right|.\ee

\begin{figure*}[t!]
  \epsfig{file=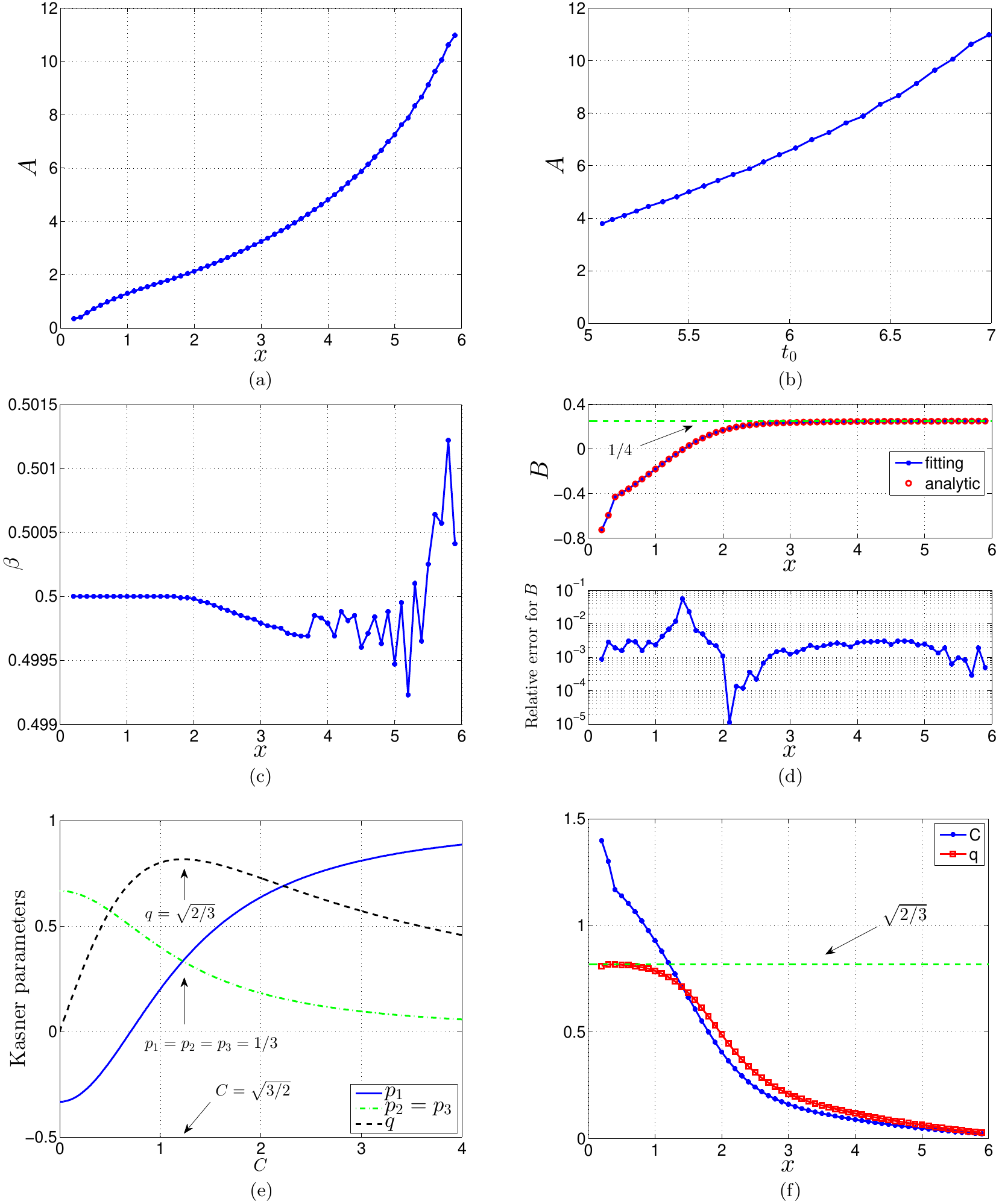, width=16cm,height=19.2cm}
  \caption{Variations of some parameters for the metric components and scalar field along the singularity curve, $r=0$, which is plotted in Fig.~\ref{fig:apparent_horizon}(a). The results are obtained by fitting the numerical results to the corresponding equations.
  (a) $A$ for Eq.~(\ref{r_asymptotic}). Namely, $r\approx A\xi^{\beta}$.
  (b) $A$ vs $t_0$, where $t_0$ is the coordinate time on the singularity curve. Currently, an asymptotic relation between $A$ and $t_0$ is unavailable.
  (c) $\beta$ for Eq.~(\ref{r_asymptotic}). Namely, $r\approx A\xi^{\beta}$.
  (d) $B$ for Eq.~(\ref{sigma_asymptotic}). Namely, $\sigma\approx B\log\xi$.
  (e) The Kasner exponents and the parameter $q$ described by Eqs.~(\ref{kasner_parameters}) and (\ref{kasner_q}), respectively. The numbers $p_1$, $p_2$, $p_3$, and $q$ are subject to Eq.~(\ref{Kasner_solution}). Namely, $p_1+p_2+p_3=1$ and $p_1^2+p_2^2+p_3^2=1-q^2$.
  (f) $C$ for Eq.~(\ref{phi_kasner}). Namely, $\phi\approx C\log \xi$. $q$ is obtained from Eq.~(\ref{kasner_q}). Namely, $q=4C/(3+2C^2)$.
  In the range of $0.2\le x\le 3.7$, 20 iterations of mesh refinement are implemented. For $x>3.7$, more iterations are added in order to make the fitting converge quickly. In the ranges of $3.8\le x\le 5.5$ and $5.6\le x\le 5.9$, 22 and 25 iterations of mesh refinement are implemented, respectively.
  }
  \label{fig:variation_singularity_curve}
\end{figure*}

We study variations of the parameters $A$, $\beta$, $B$, and $C$ present in Eqs.~(\ref{r_asymptotic})-(\ref{phi_asymptotic}) along the singularity curve by fitting the numerical results to corresponding analytic expressions. The results are plotted in Fig.~\ref{fig:variation_singularity_curve}. The results imply that at places far away from the center $x=0$, the contribution from the scalar field $\phi$ is negligible, and the spacetime is very similar to the one of a Schwarzschild black hole in Kruskal coordinates. Equation~(\ref{r_Kruskal}) reveals that $A=4mt_{0}^{1/2}$ and $\beta=1/2$ for a Schwarzschild black hole. In the collapse case, we plot the relation of $A$ vs $t_0$ in Fig.~\ref{fig:variation_singularity_curve}(b), while an approximate analytic expression for $A$ vs $t_0$ is unavailable yet. Figure~\ref{fig:variation_singularity_curve}(c) shows that $\beta$ is very close to $1/2$. In Fig.~\ref{fig:variation_singularity_curve}(d), we plot results for $B$ obtained both via fitting the numerical results and the analytic expression $B=(1-2C^2)/4$ [see Eq.~(\ref{sigma_asymptotic_2})]. We also compute the relative errors between the two sets of results. The results from the two approaches are very close. They asymptote to $1/4$ at places far from the center $x=0$. This is consistent with the Schwarzschild black hole case, in which $B=1/4$.

As functions of $C$, the Kasner exponents and the parameter $q$ are plotted in Fig.~\ref{fig:variation_singularity_curve}(e). Equation~(\ref{Kasner_solution}) constrains the parameter $q$ as $q\le\sqrt{2/3}$. This is verified in Fig.~\ref{fig:variation_singularity_curve}(e). When $C=\sqrt{3/2}\approx 1.22$, there are $q=\sqrt{2/3}\approx 0.82$ and $p_1=p_2=p_3=1/3$. By fitting the numerical results to Eq.~(\ref{phi_asymptotic}), we obtain variations of $C$ and $q$ along the singularity curve, as plotted in Fig.~\ref{fig:variation_singularity_curve}(f). In the direction from $x=5.5$ toward $x=0$, $q$ increases and approaches the maximum value, $\sqrt{2/3}$, near $x=0.3$. Note that $q$ describes the contribution of the scalar field $\phi$. The variation of $q$ can be interpreted in a straightforward way. During the collapse, $\psi$ and $\phi$ move toward the center $x=0$. Due to interactions between the scalar fields and spacetime, the major energy of $\phi$ arrives at the formed singularity near $x=0.3$, and contributes most at this point.

One may wonder what the asymptotic values for $A$, $\beta$, $B$, and $C$ are as $x$ approaches zero along the singularity curve. Another issue is the running of these parameters with respect to the scale of $\xi$. Letting the spatial coordinate $x$ take a fixed value, we implement mesh refinement with different iterations. Correspondingly, $\xi$ reaches different scales. We obtain $C$ by fitting the numerical results to Eq.~(\ref{phi_asymptotic}). We find that $C$ is running with respect to the scale of $\xi$. For example, at $x=2.5$, $C$ decreases about three percent when the scale of $\xi$ is reduced from $10^{-3}$ to $10^{-8}$. However, detailed studies of such issues are beyond the scope of this paper.

In the Einstein frame where we are working, the gravitational theory is similar to general relativity. Two scalar fields, $\phi$ (or $f'$) and $\psi$, are present. However, near the singularity, the contributions to the spacetime from the physical scalar field $\psi$ and the potential for $\phi$ are negligible. The field $\phi$ is almost massless. The contribution from the almost-massless field $\phi$ is important. Therefore, this case is essentially the same as single massless scalar (spherical) collapse in general relativity. The Kasner solution we obtained for spherical scalar collapse in $f(R)$ theory is also the corresponding Kasner solution for single massless scalar collapse in general relativity.

The statement that spherical collapse in general relativity can end up with a Schwarzschild black hole has been verified by various numerical simulations. Hawking showed that stationary black holes as the final states of Brans-Dicke collapses are also the solutions of general relativity~\cite{Hawking}.
This conclusion has been numerically confirmed in Refs.~\cite{Scheel_1,Scheel_2,Shibata}. A static black hole in scalar-tensor theories [including $f(R)$ theory] has a de Sitter-Schwarzschild solution. In the $f(R)$ theory case, $f'$ would stay at the minimum of the potential, $U(f')$. However, numerical simulations show that in the collapse process, $f'$ crosses the minimum of the potential, and 
\break
\break
\break
asymptotes to zero as the singularity is approached. 
Namely, the static and dynamical solutions are considerably different. One may wonder whether the collapse will lead to the static solution eventually. Preliminary explorations show that this may not be a trivial question. Further explorations of this problem are omitted in this paper.

\section{View from the Jordan frame\label{sec:view_JF}}

Originally, $f(R)$ gravity is defined in the Jordan frame. For computational convenience, we transform $f(R)$ gravity from the Jordan frame into the Einstein frame. After the results have been obtained in the Einstein frame, we convert these results back into the Jordan frame in this section. We examine the Ricci scalar, Weyl scalar, Weyl tensor, and Kasner solution in the Jordan frame.

\subsection{Ricci scalar}
\begin{figure*}[t!]
  \epsfig{file=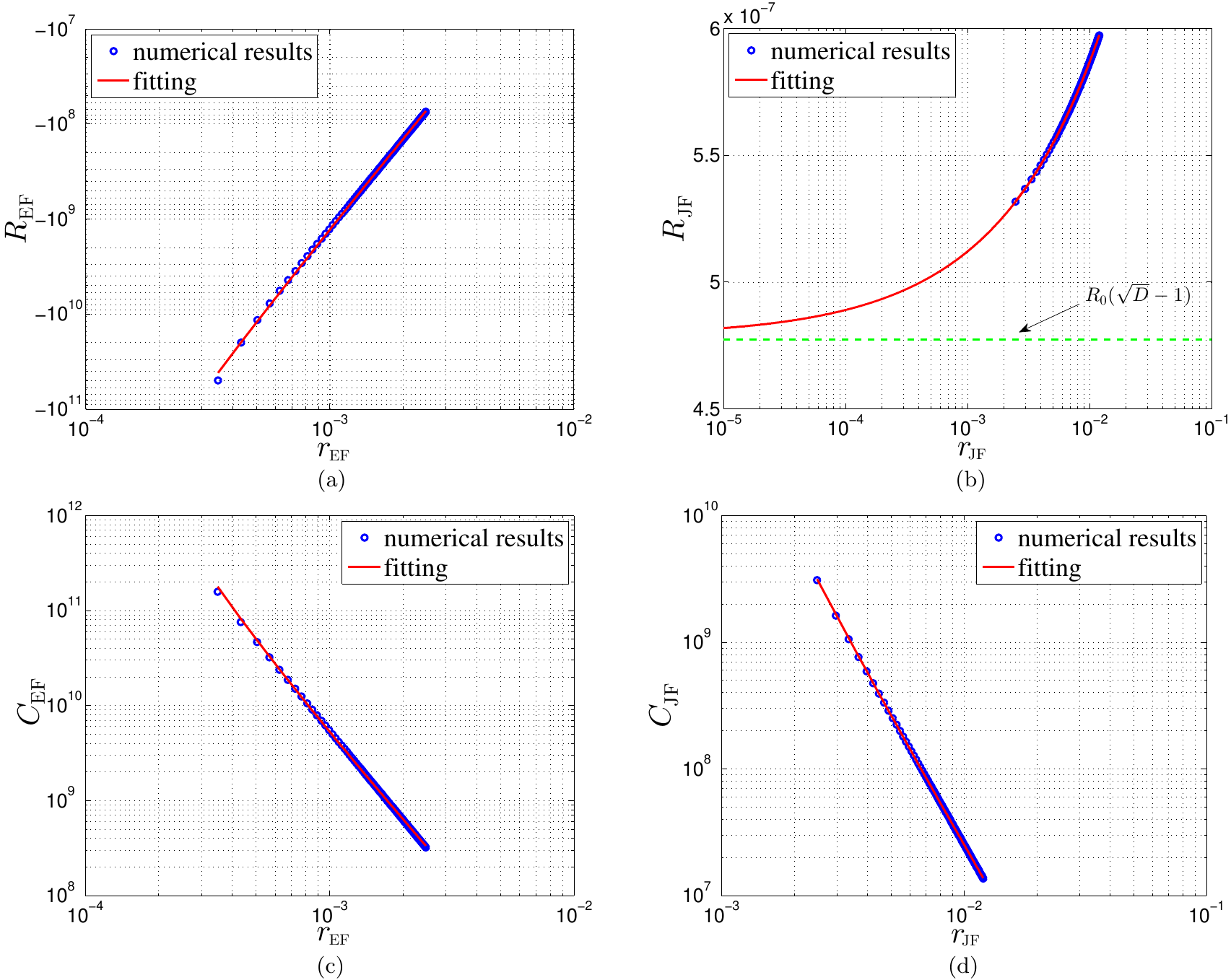, width=17cm,height=13.6cm}
  \caption{Curvature invariants near the singularity in the Einstein and Jordan frames in the collapse of the Hu-Sawicki model expressed by Eq.~(\ref{f_R_Hu_Sawicki}). The results are for $(x=2.5,t=t)$. We fit the results as follows.
  (a) $\log |R_{\mbox{\scriptsize EF}}|=a+b\log (r_{\mbox{\scriptsize EF}}+c)$. We fix $a$ to $-0.592$, which is the modified analytic value for $a$. $b=-3.1102\pm 0.0009$, $c=(-3.0\pm 0.2)\times 10^{-6}$. $R_{\mbox{\scriptsize EF}}$ diverges in the vicinity of the singularity, due to contributions from the scalar field, $\phi[\equiv(\sqrt{3/2}\log f')/\kappa]$.
  (b) $R_{\mbox{\scriptsize JF}}=a+b(r_{\mbox{\scriptsize JF}})^{c}$, $a=(4.7869\pm0.0001)\times 10^{-7}$, $b=(1.1294\pm0.0002)\times 10^{-6}$, $c=0.50983\pm0.00006$. As shown in Fig.~\ref{fig:Kasner_metric_fields}(c), when the singularity is approached, $f'$ asymptotes to zero. Consequently, with Eq.~(\ref{R_JF_kasner}), $R_{\mbox{\scriptsize JF}}$ will approach to a constant: $R_{0}(\sqrt{D}-1)$.
  (c) $\log C_{\mbox{\scriptsize EF}}=a+b\log (r_{\mbox{\scriptsize EF}}+c)$. We fix $a$ to $2.072$, which is the modified analytic value for $a$.
  $b=-2.9117\pm 0.0008$, $c=(7.0\pm 0.2)\times 10^{-6}$.
  (d) $\log C_{\mbox{\scriptsize JF}}=a+b\log (r_{\mbox{\scriptsize JF}}+c)$, $a=1.93\pm 0.03$, $b=-3.265\pm 0.007$, $c=(2.6\pm 0.1)\times 10^{-4}$.
  }
  \label{fig:invariants}
\end{figure*}

First, using the asymptotic expressions for $r$ (\ref{r_asymptotic}) and $\sigma$ (\ref{sigma_asymptotic_2}), we compute the Ricci scalar in the Einstein frame as follows:
\begin{eqnarray}
R_{\mbox{\scriptsize EF}} & = & 2e^{2\sigma}\bigg[ -\sigma_{,tt}+\sigma_{,xx}+\frac{2(r_{,tt}-r_{,xx})}{r}\nonumber\\
 && \hphantom{158pt}+\frac{(r_{,t})^{2}-(r_{,x})^2}{r^2}+\frac{e^{-2\sigma}}{r^2}\bigg]\nonumber\\
 &\approx& (J^2-1)\cdot C^2\cdot\xi^{-\frac{(3+2C^2)}{2}}\nonumber\\
 &\approx& (J^2-1)\cdot C^2\cdot \left(\frac{r_{\mbox{\scriptsize EF}}}{A}\right)^{-3-2C^2},
\end{eqnarray}
where $J$ is the slope of the singularity curve, and
\be J^2\approx \frac{r_{,xx}}{r_{,tt}}\approx \frac{\sigma_{,xx}}{\sigma_{,tt}}\approx \frac{(r_{,x})^2}{(r_{,t})^2}.\ee
Refer to arguments in Sec.~\ref{sec:results_dynamics} for details on the above equation. We use $r$ and $r_{\mbox{\scriptsize EF}}$ to denote the quantity $r$ in the Einstein frame and $r_{\mbox{\scriptsize JF}}$ in the Jordan frame. The numerical results and fitting results for $R_{\mbox{\scriptsize EF}}$ in the vicinity of the singularity on the slice $(x=2.5,t=t)$ are plotted in Fig.~\ref{fig:invariants}(a). We fit the numerical results
according to $\log |R_{\mbox{\scriptsize EF}}|=a+b\log (r_{\mbox{\scriptsize EF}}+c)$. We fix $a$ to $-0.592$, which is the modified analytic value for $a$ as discussed below. The fitting results are
\be
\begin{array}{l l}
  b=-3.1102\pm 0.0009,\\
  \\
  c=(-3.0\pm 0.2)\times 10^{-6}.
\end{array}
\nonumber\ee
The analytic results are
\be
\begin{array}{l l}
  a_{\mbox{\scriptsize analytic}}=\log\left[(1-J^2)C^{2}A^{3+2C^2}\right]=0.0922\pm 0.0003,\\
  \\
  b_{\mbox{\scriptsize analytic}}=-3-2C^2=-3.11587\pm 0.00003,\\
  \\
  c_{\mbox{\scriptsize analytic}}=0.
\end{array}
\nonumber\ee

In the above computations, we have used the approximate expression for $\sigma$ (\ref{sigma_asymptotic}), $\sigma\approx B\log\xi$. This expression is valid when $r$ is close enough to zero. The fitting results for $\sigma$ for the slice $(x=2.5,t=t)$ are $\sigma=-0.34224+0.22108\log \xi$. If we used this more accurate expression, the modified analytic value for $a$
would be $a_{\mbox{\scriptsize analytic-modify}}=-0.592\pm 0.001$.

The Ricci scalar in the Jordan frame for the Hu-Sawicki model can be obtained from Eq.~(\ref{R_Hu_Sawicki}). In the vicinity of the singularity, $f'\ll 1$. Then Eq.~(\ref{R_Hu_Sawicki}) becomes
\begin{eqnarray}
R_{\mbox{\scriptsize JF}} & = & R_{0}\left[\sqrt{\frac{D}{1-f'}}-1\right] \nonumber\\
  & \approx & R_{0}(\sqrt{D}-1) + \frac{R_{0}\sqrt{D}}{2}\left(\frac{r_{\mbox{\scriptsize JF}}}{A}\right)^{\frac{2\sqrt{2/3}C}{1-\sqrt{2/3}C}},
\label{R_JF_kasner}
\end{eqnarray}
where we have used
\be f'\equiv \chi=e^{\sqrt{\frac{2}{3}}\phi}\approx \xi^{\sqrt{\frac{2}{3}}C},\label{f_prime_kasner}\ee
\be r_{\mbox{\scriptsize JF}}=r_{\mbox{\scriptsize EF}}\cdot\chi^{-\frac{1}{2}}\approx A\xi^{\frac{1-\sqrt{2/3}C}{2}}.\ee
Note that in this paper we have set $8\pi G=\kappa^2=1$. Equation (\ref{R_JF_kasner}) reveals that as $r_{\mbox{\scriptsize JF}}$ asymptotes to zero, $R_{\mbox{\scriptsize JF}}$ will approach a constant: $R_{0}(\sqrt{D}-1)$. The numerical results and fitting results for $R_{\mbox{\scriptsize JF}}$ are shown in Fig.~\ref{fig:invariants}(b). The numerical results are fit according to
$R_{\mbox{\scriptsize JF}}=a+b(r_{\mbox{\scriptsize JF}})^{c}$. The fitting results are
\be
\begin{array}{l l}
  a=(4.7869\pm0.0001)\times 10^{-7},\\
  \\
  b=(1.1294\pm0.0002)\times 10^{-6},\\
  \\
  c=0.50983\pm0.00006.
\end{array}
\nonumber\ee
The corresponding analytic results are
\be
\begin{array}{l l}
  a_{\mbox{\scriptsize analytic}}=R_{0}(\sqrt{D}-1)=4.77\times 10^{-7},\\
  \\
  b_{\mbox{\scriptsize analytic}}=\frac{R_{0}\sqrt{D}}{2}A^{-\frac{2\sqrt{2/3}C}{1-\sqrt{2/3}C}}=(1.7014\pm0.0001)\times 10^{-6},\\
  \\
  c_{\mbox{\scriptsize analytic}}=\frac{2\sqrt{2/3}C}{1-\sqrt{2/3}C}=0.48920\pm0.00008.
\end{array}
\nonumber\ee
In the above computations, we have used the approximate expression for $\phi$ (\ref{phi_asymptotic}), $\phi\approx C\log\xi$. This expression is valid when $r$ is close enough to zero. The fitting results for $\phi$ for the slice $(x=2.5,t=t)$ are $\phi=-0.5118+0.2407\log \xi$ [see Fig.~\ref{fig:Kasner_fit}(a)]. If we used this more accurate expression, the modified analytic value for $b$ would be $b_{\mbox{\scriptsize analytic-modify}}=(1.0114\pm0.0004)\times 10^{-6}$.
\subsection{Weyl scalar}
The Ricci tensor and Ricci scalar include information on the traces of the Riemann tensor, while the trace-free parts are described by the Weyl tensor and Weyl scalar. We consider the Weyl scalar and Weyl tensor in this and the next subsections, respectively. It is convenient to define
\be A_{\scriptsize W}=\sigma_{,xx}-\sigma_{,tt}+\frac{r_{,xx}-r_{,tt}}{r}+\frac{(r_{,t})^{2}-(r_{,x})^2}{r^2}+\frac{e^{-2\sigma}}{r^2}.\ee
Then in the Einstein frame, the Weyl scalar is
\begin{eqnarray}
C_{\mbox{\scriptsize EF}} & \equiv & \sqrt{C_{\alpha\beta\mu\nu}C^{\alpha\beta\mu\nu}}=\sqrt{\frac{4}{3}}e^{2\sigma}A_{\scriptsize W}\nonumber\\
 &\approx& \frac{3-2C^2}{2\sqrt{3}}(1-J^2)\xi^{-\frac{3+2C^2}{2}}\nonumber\\
 &\approx& \frac{3-2C^2}{2\sqrt{3}}(1-J^2)\left(\frac{r_{\mbox{\scriptsize EF}}}{A}\right)^{-3-2C^2},
\end{eqnarray}
where $C_{\alpha\beta\mu\nu}$ is the Weyl tensor. The numerical and fitting results for $C_{\mbox{\scriptsize EF}}$ are plotted in Fig.~\ref{fig:invariants}(c). We fit the numerical results according to
$\log C_{\mbox{\scriptsize EF}}=a+b\log (r_{\mbox{\scriptsize EF}}+c)$. We fix $a$ to $2.072$, which is the modified analytic value for $a$ as discussed below. The results are
\be
\begin{array}{l l}
  b=-2.9117\pm 0.0008,\\
  \\
  c=(7.0\pm 0.2)\times 10^{-6}.
\end{array}
\nonumber\ee
The analytic results are
\be
\begin{array}{l l}
\begin{split}
  a_{\mbox{\scriptsize analytic}}&=\log\left[\frac{3-2C^2}{2\sqrt{3}}(1-J^2)A^{3+2C^2}\right]\\
                                 &=2.7574\pm 0.0001,\end{split}\\
  \\
  b_{\mbox{\scriptsize analytic}}=-3-2C^2=-3.11587\pm 0.00003,\\
  \\
  c_{\mbox{\scriptsize analytic}}=0.
\end{array}
\nonumber\ee
If we used the more accurate expression for $\sigma$, $\sigma=-0.34224+0.22108\log \xi$, the modified analytic value for $a$ would be
$a_{\mbox{\scriptsize analytic-modify}}=2.072\pm0.001$.

The Weyl scalar in the Jordan frame is \cite{Wald_1984}
\begin{eqnarray}
C_{\mbox{\scriptsize JF}} &=& f'\cdot C_{\mbox{\scriptsize EF}}\nonumber\\
 &\approx& \frac{3-2C^2}{2\sqrt{3}}(1-J^2)\left(\frac{r_{\mbox{\scriptsize JF}}}{A}\right)^{-\frac{3+2C^2-2\sqrt{2/3}C}{1-\sqrt{2/3}C}}.
\end{eqnarray}
We fit the numerical results according to $\log C_{\mbox{\scriptsize JF}}=a+b\log (r_{\mbox{\scriptsize JF}}+c)$. The results are
\be
\begin{array}{l l}
  a=1.93\pm 0.03,\\
  \\
  b=-3.265\pm 0.007,\\
  \\
  c=(2.6\pm 0.1)\times 10^{-4}.
\end{array}
\nonumber\ee
The analytic results are
\be
\begin{array}{l l}
  a_{\mbox{\scriptsize analytic}}=3.0230\pm 0.0001,\\
  \\
  b_{\mbox{\scriptsize analytic}}=-3.3888\pm 0.0001,\\
  \\
  c_{\mbox{\scriptsize analytic}}=0.
\end{array}
\nonumber\ee
If we used the more accurate expressions for $\phi$ and $\sigma$, $\phi=-0.5118+0.2407\log\xi$, and $\sigma=-0.34224+0.22108\log \xi$, the modified analytic value for $a$ would be $a_{\mbox{\scriptsize analytic-modify}}=2.6247\pm0.0002$.

\subsection{Weyl tensor}
The Weyl tensor in the format of $C^{\alpha}_{\beta\mu\nu}$ is invariant under conformal transformations. We compute one component of the Weyl tensor,
\be
\resizebox{0.88\hsize}{!}{$C^{t}_{xtx} = \frac{1}{3}A_{\scriptsize W}
\approx \frac{1}{3}\left[(1-J^2)\frac{3-2C^2}{4}\xi^{-2}+A^{-2}\xi^{-\frac{3-2C^2}{2}}\right].$}
\label{weyl_tensor}
\ee
We also compute the metric components in the Jordan frame in the vicinity of the singularity curve using the transformation relation, $g_{\mu\nu}^{(\mbox{\scriptsize EF})}=\chi\cdot g_{\mu\nu}^{(\mbox{\scriptsize JF})}$:
\be r_{\mbox{\scriptsize JF}}=r_{\mbox{\scriptsize EF}}\cdot\chi^{-\frac{1}{2}}\approx A\xi^{\frac{1-\sqrt{2/3}C}{2}},\label{r_JF}\ee
\be \left. e^{-\sigma}\right|_{\mbox{\scriptsize JF}}=\left. e^{-\sigma}\right|_{\mbox{\scriptsize EF}}\cdot\chi^{-\frac{1}{2}}
\approx \xi^{\frac{-1+2C^2-2\sqrt{2/3}C}{4}}.\label{sigma_JF}\ee

Equations (\ref{weyl_tensor})--(\ref{sigma_JF}) show that $C=\sqrt{3/2}$ is a special point. As $\xi$ approaches zero, when $0<C<\sqrt{3/2}$, $C^{t}_{xtx}$ and $\left. e^{-\sigma}\right|_{\mbox{\scriptsize JF}}$ become positive infinity, and $r_{\mbox{\scriptsize JF}}$ asymptotes to zero. However, when $C>\sqrt{3/2}$, $C^{t}_{xtx}$ becomes negative infinity, $r_{\mbox{\scriptsize JF}}$ becomes positive infinity, and $\left. e^{-\sigma}\right|_{\mbox{\scriptsize JF}}$ asymptotes to zero. Further explorations of these issues are beyond the scope of this paper. Since the Weyl tensor is invariant under conformal transformations, $C^{t}_{xtx}$ will also become positive infinity in the Jordan frame in the case of $0<C<\sqrt{3/2}$. Moreover, the radius of the apparent horizon for the black hole in the Jordan frame can be obtained from Eq.~(\ref{r_JF}). Consequently, a black hole can also be formed in the Jordan frame. The scalar degree of freedom $f'$ will approach zero as $r_{\mbox{\scriptsize JF}}$ asymptotes to zero.

\subsection{Kasner solution in the Jordan frame}
In the Jordan frame, the proper time for the case of $0<C<\sqrt{3/2}$ is
\be
\resizebox{1\hsize}{!}{$\tau_{\mbox {\scriptsize JF}}=\int_0^{\xi}\left. e^{-\sigma}\right|_{\mbox {\scriptsize JF}} d\xi
\approx\frac{4}{3+2C^2-2\sqrt{2/3}C}\xi^{\frac{3+2C^2-2\sqrt{2/3}C}{4}}.$}
\ee
Therefore, $r_{\mbox{\scriptsize JF}}$, $\left. e^{-\sigma}\right|_{\mbox{\scriptsize JF}}$, and $\phi$ can be written in terms of $\tau_{\mbox{\scriptsize JF}}$ as follows:
\be r_{\mbox{\scriptsize JF}}\approx A\left(\frac{3+2C^2-2\sqrt{2/3}C}{4}
\tau_{\mbox{\scriptsize JF}}\right)^{\frac{2\left(1-\sqrt{2/3}C\right)}{3+2C^2-2\sqrt{2/3}C}},
\label{r_kasner_JF}\ee
\be \left. e^{-\sigma}\right|_{\mbox{\scriptsize JF}}
\approx \left(\frac{3+2C^2-2\sqrt{2/3}C}{4}\tau_{\mbox{\scriptsize JF}}\right)^{\frac{-1+2C^2-2\sqrt{2/3}C}{3+2C^2-2\sqrt{2/3}C}},
\label{sigma_kasner_JF}\ee
\be \phi\approx \frac{4C}{3+2C^2-2\sqrt{2/3}C}\log \tau_{\mbox{\scriptsize JF}}.\label{phi_kasner_JF}\ee
Comparing Eqs.~(\ref{r_kasner_JF})-(\ref{phi_kasner_JF}) to (\ref{Kasner_solution}), we have
\be ^{(\mbox{\scriptsize JF})}p_1=\frac{-1+2C^2-2\sqrt{2/3}C}{3+2C^2-2\sqrt{2/3}C}, \ee
\be ^{(\mbox{\scriptsize JF})}p_2=\mspace{1mu}^{(\mbox{\scriptsize JF})}p_3=\frac{2(1-\sqrt{2/3}C)}{3+2C^2-2\sqrt{2/3}C},\ee
\be ^{(\mbox{\scriptsize JF})}q=\frac{4C}{3+2C^2-2\sqrt{2/3}C}.\label{kasner_q_JF}\ee
Obviously, $^{(\mbox{\scriptsize JF})}p_{1}$, $^{(\mbox{\scriptsize JF})}p_{2}$, $^{(\mbox{\scriptsize JF})}p_{3}$, and $^{(\mbox{\scriptsize JF})}q$ do not satisfy $^{(\mbox{\scriptsize JF})}p_{1}+ {^{(\mbox{\scriptsize JF})}p_{2}}+ {^{(\mbox{\scriptsize JF})}p_{3}}=1$ and
$^{(\mbox{\scriptsize JF})}{p_{1}}^{2}+ {^{(\mbox{\scriptsize JF})}{p_{2}}^{2}}+ {^{(\mbox{\scriptsize JF})}{p_{3}}^{2}}
=1- {^{(\mbox{\scriptsize JF})}q^2}$. This is because in the Jordan frame, the scalar degree of freedom, $f'$, is not minimally coupled to gravity, while that is the case in the Einstein frame or general relativity.

\begin{figure}
  \epsfig{file=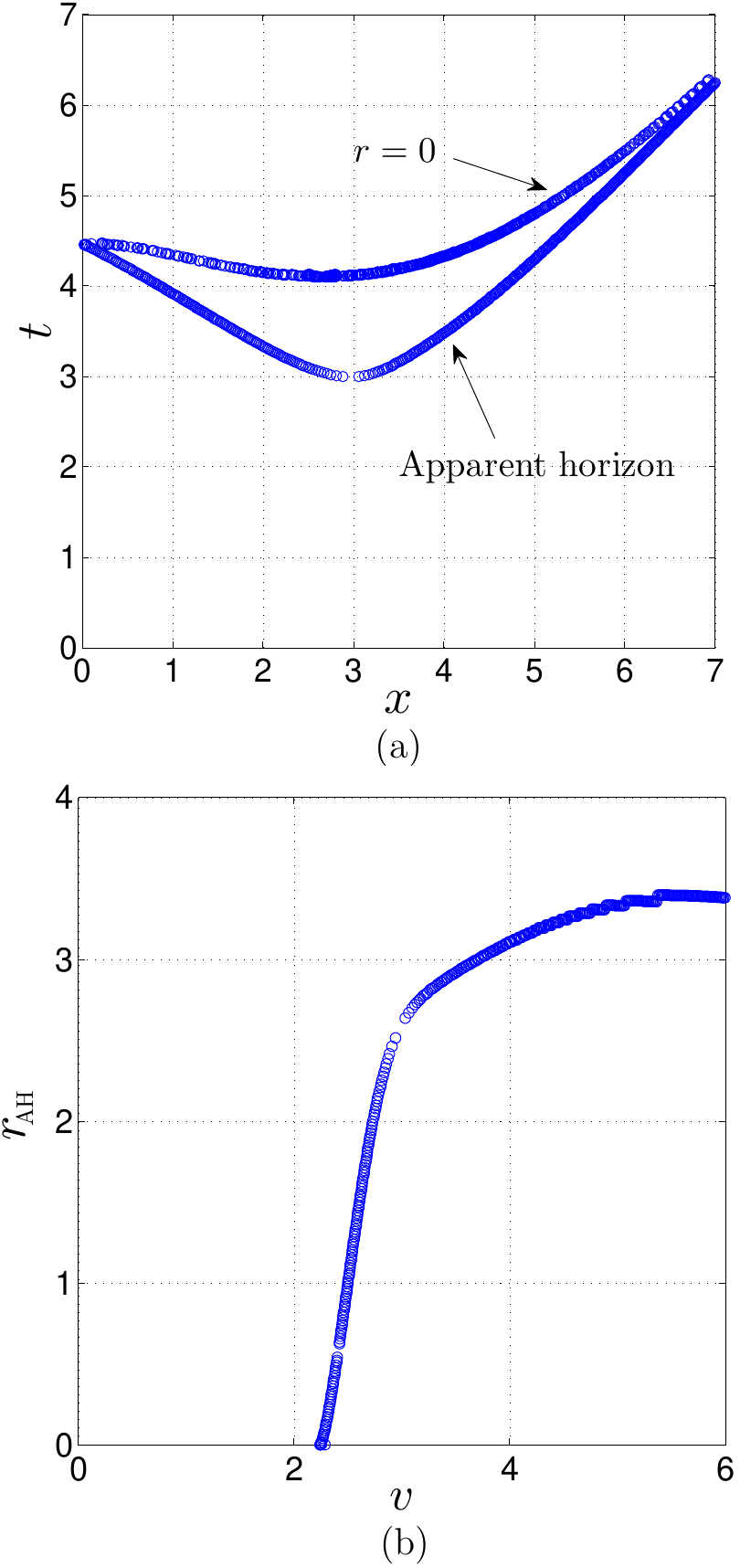, width=6.6cm,height=14.18cm}
  \caption{Apparent horizon of the black hole obtained from spherical collapse for the Hu-Sawicki model described by Eq.~(\ref{f_R_Hu_Sawicki}), with $D=1.05$ and $R_{0}=5\times10^{-6}$. $v=(x+t)/2$.}
  \label{fig:apparent_horizon_variation}
\end{figure}

\begin{figure*}[t!]
  \epsfig{file=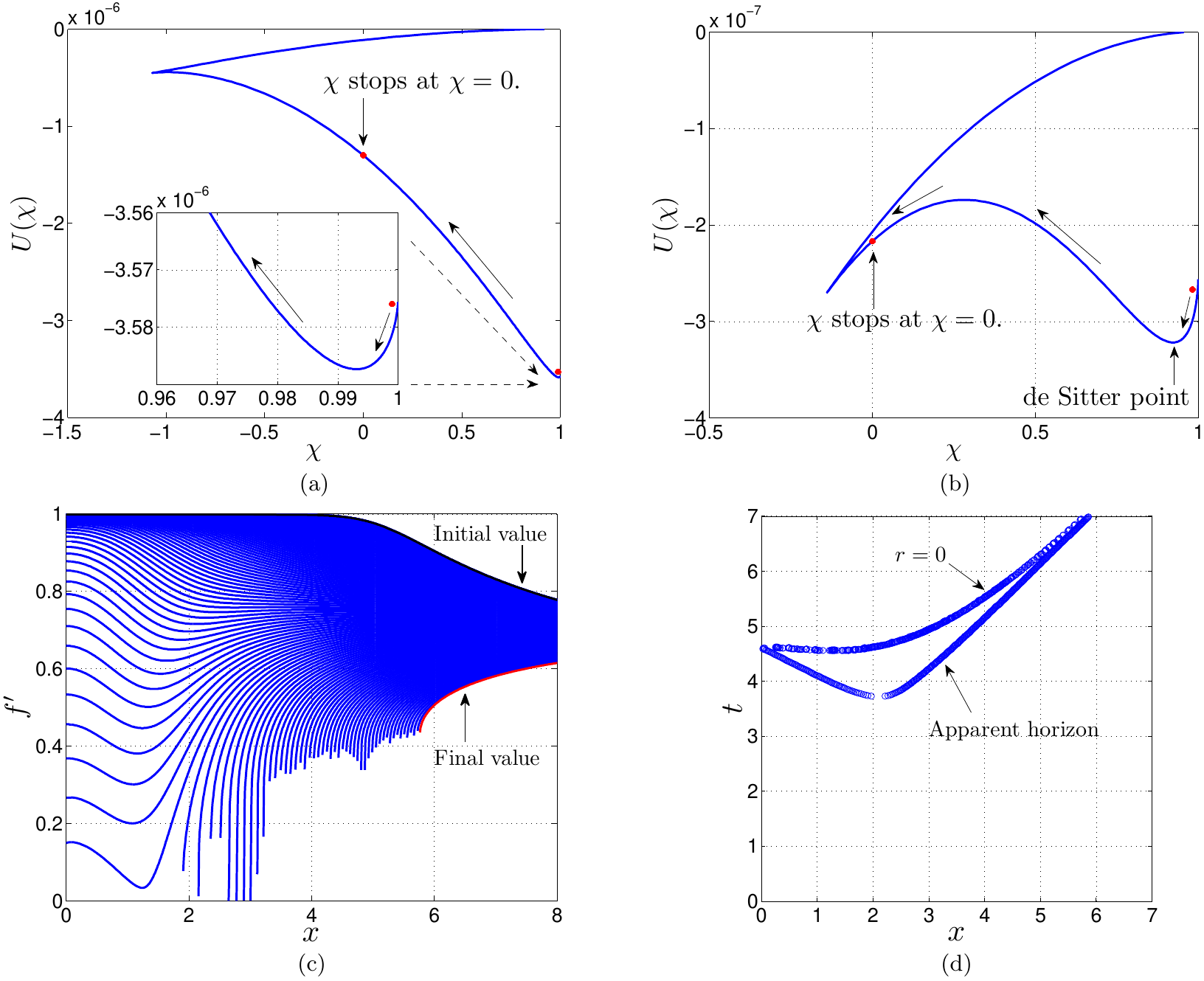, width=17cm,height=13.6cm}
  \caption{Spherical collapse in the Starobinsky model (\ref{f_R_Starobinsky}), with $n=2$ and $R_{0}=5\times 10^{-6}$. (a) is for $D=2$, and (b)-(d) for $D=1.1$. The results are similar to those in the Hu-Sawicki model. Since the potential is not important in the vicinity of the singularity, $\chi(\equiv f')$ asymptotes to zero as the singularity is approached, no matter what the potential looks like.}
  \label{fig:Starobinsky}
\end{figure*}

\section{Collapses in more general models\label{sec:collapse_general}}

We have studied spherical collapse for one of the simplest versions of the Hu-Sawicki model in the Einstein frame. In this section, we will discuss collapses in more general cases. We will examine how the parameter $D$ in the Hu-Sawicki model~(\ref{f_R_Hu_Sawicki}) affects the results. Spherical collapse for another typical dark energy model, the Starobinsky model, will be explored.

\subsection{Collapse for the Hu-Sawicki model in general cases}

In one of the simplest versions of the Hu-Sawicki model, described by Eq.~(\ref{f_R_Hu_Sawicki}), the parameter $D$ is set to $1.2$. Now we let $D$ take a smaller value $1.05$. This means that the dark energy will play a less important role. The results in this configuration are plotted in Fig.~\ref{fig:apparent_horizon_variation}. Not surprisingly, in comparison to Fig.~\ref{fig:apparent_horizon} with $D=1.2$, in this new case, it takes less time to form a black hole, and the radius of the apparent horizon of the formed black hole is larger. In the case of $D=1.2$, the apparent horizon starts to form at $t=3.6$, and the radius of the black hole is about $2.2$. In the case of $D=1.05$, the apparent horizon starts to form at $t=3.0$, and the radius of the black hole is about $3.4$.

\subsection{Collapse for the Starobinsky model}

We consider spherical collapse for the Starobinsky model, which can be expressed as follows~\cite{Starobinsky_2}:
\be f(R)=R+DR_{0}\left[\left(1+\frac{R^2}{{R_0}^2}\right)^{-n}-1\right],\label{f_R_Starobinsky}\ee
where $D$ and $n$ are positive parameters, and $R_0$ has the same order of magnitude as the currently observed effective cosmological constant. In this paper, we set $R_0$ to $5\times 10^{-6}$.

We simulate collapses with $n=1$ and $n=2$. Note that the case of $n=1$ for the Starobinsky model (\ref{f_R_Starobinsky}) is identical to the case of $n=2$ for the Hu-Sawicki model (\ref{f_R_Hu_Sawicki_general}). The results with $n=1$ and $n=2$ for the Starobinsky model are similar, and we only present results of the $n=2$ case in Fig.~\ref{fig:Starobinsky}. The potentials in Figs.~\ref{fig:Starobinsky}(a) and (b) are for $D=2$ and $D=1.1$, respectively.   The results of these two cases are also similar, and only those for $D=1.1$ are plotted in Figs.~\ref{fig:Starobinsky}(c) and (d). These results are close to those for the Hu-Sawicki model. Since the potential is not important in the vicinity of the singularity, $f'$ asymptotes to zero as the singularity is approached, no matter what the potential looks like near $f'=0$. [see Figs.~\ref{fig:Starobinsky}(a) and (b)].

\section{Conclusions\label{sec:conclusions}}
Spherical scalar collapse in $f(R)$ gravity was simulated in this paper. A black hole formation was obtained. The dynamics of the metric components, the scalar degree of freedom $f'$, and a physical scalar field during the collapse process, including near the singularity, were studied. The results confirmed the BKL conjecture.

Originally, $f(R)$ gravity was built in the Jordan frame. For computational convenience, we transformed $f(R)$ gravity from the Jordan frame into the Einstein frame, in which the gravitational theory is similar to general relativity. The double-null coordinates were employed. These coordinates enabled us to study the dynamics both inside and outside of the horizon of the formed black hole. Two typical dark energy $f(R)$ models, the Hu-Sawicki model and Starobinsky model, were taken as example models in this paper. Mesh refinement and asymptotic analysis were applied to study the dynamics in the vicinity of the singularity of the formed black hole.

The dark energy $f(R)$ theory is a modification of general relativity at low curvature scale. Inside a sphere whose matter density is much greater than the dark energy density and whose radius is large enough, $f'$ is coupled to the matter density and is close to $1$. Accordingly, $f(R)$ gravity is reduced to general relativity and the modification term is negligible. However, during the collapse, the matter moves to the center of the scalar sphere, which forms a black hole at a later stage. Then, $f'$ loses the coupling and becomes almost massless. Due to the strong gravity from the singularity and the low mass of $f'$, $f'$ crosses its de Sitter value and asymptotes to zero as the singularity is approached. Simultaneously, the modification term in the function $f(R)$ takes effect and even becomes dominant. Therefore, the solution of the dynamical collapse is significantly different from the static solution---it is not the de Sitter-Schwarzschild solution.

Near the singularity, in the equations of motion for the metric components and the scalar fields, the metric component terms are more important than the scalar field ones. The field $\phi$, transformed from the scalar degree of freedom $f'$, dominates the competition between $\phi$ and the physical field $\psi$. The field $\phi$ contributes more to the dynamics of the metric components than $\psi$ does. In the equations of motion for the metric components and $\phi$, the contributions of $\psi$ are negligible. However, the effect of $\phi$ on the evolution of $\psi$ is visible. The field $\phi$ or effective dark energy tries to stop the collapse of $\psi$. The metric components and the scalar field $\phi$ are described by the Kasner solution. These results supported the BKL conjecture well.

In the vicinity of the singularity, the field $\psi$ can be omitted. The field $\phi$ remains, with the potential being negligible. Therefore, the Kasner solution for spherical scalar collapse in $f(R)$ theory that we obtained is also the Kasner solution for spherical scalar collapse in general relativity.

In studies of cosmological dynamics and local tests of $f(R)$ theory, much attention has been given to the right side and the minimum area of the potential as plotted in Fig.~\ref{fig:potential_Hu_Sawicki}~\cite{Frolov_2}. In the early Universe, the scalar field $f'$ is coupled to the matter density and is close to $1$. In the later evolution, $f'$ goes down toward the minimum of the potential, oscillates, and eventually stops at the minimum. In the oscillation epoch, $f'$ does not deviate too far from the minimum. However, in the collapse process toward a black hole formation, the strong gravity from the black hole pulls $f'$ in the left direction to a place far away from the minimum. Consequently, the left side of the potential needs more care in the collapse problem.

\section*{Acknowledgments}
This  work  was  supported  by  the  Discovery  Grants  program  of  the  Natural  Sciences  and Engineering Research Council of Canada. The authors would  like  to thank Matthew W. Choptuik, Tony Chu,  Mariusz P. Dabrowski, Levon Pogosian, and Howard Trottier for useful discussions.  The authors also thank the referee for helpful comments. J.Q.G. thanks the participants for helpful discussions when a seminar on this work was given at the Tata Institute of Fundamental Research, Mumbai, India.

\begin{appendix}
\section{Einstein tensor and Energy-momentum tensor of a massive scalar field\label{sec:appendix_A}}
In this Appendix, we give specific expressions of the Einstein tensor and energy-momentum tensor of a massive scalar field. In double-null coordinates (\ref{double_null_metric}), some components of the Einstein tensor can be expressed as follows:
\be
\begin{split}
G^{t}_{t}=&\frac{2e^{2\sigma}}{r^2}\bigg[r(r_{,t}\sigma_{,t}+r_{,x}\sigma_{,x})+rr_{xx}\\
         & +\frac{1}{2}(-{r_{,t}}^2+{r_{,x}}^2)-\frac{1}{2}e^{-2\sigma}\bigg],
\end{split}
\label{G_tt}
\ee

\be
\begin{split}
G^{x}_{x}=&\frac{2e^{2\sigma}}{r^2}\bigg[-r(r_{,t}\sigma_{,t}+r_{,x}\sigma_{,x})-rr_{tt}\\
          &+\frac{1}{2}(-{r_{,t}}^2+{r_{,x}}^2)-\frac{1}{2}e^{-2\sigma}\bigg],
\end{split}
\label{G_xx}\ee

\be G^{\theta}_{\theta}=G^{\phi}_{\phi}=\frac{e^{2\sigma}}{r}\left[-r_{,tt}+r_{,xx}-r(-\sigma_{,tt}+\sigma_{,xx})\right],
\label{G_theta_theta}\ee

\be G_{uu}=-\frac{2}{r}(r_{,uu}+2\sigma_{,u}r_{,u}), \ee

\be G_{vv}=-\frac{2}{r}(r_{,vv}+2\sigma_{,v}r_{,v}). \ee

For a massive scalar field with energy-momentum tensor
\be T_{\mu\nu}=\phi_{,\mu}\phi_{,\nu}-g_{\mu\nu}\left[\frac{1}{2}g^{\alpha\beta}\phi_{,\alpha}\phi_{,\beta}+V(\phi)\right],
\label{T_mu_nu}\ee
there are
\be T^{t}_{t}=-e^{2\sigma}\left[\frac{1}{2}(\phi_{,t}^2+\phi_{,x}^2)+e^{-2\sigma}V(\phi)\right],
\label{T_tt}\ee
\be T^{x}_{x}=e^{2\sigma}\left[\frac{1}{2}(\phi_{,t}^2+\phi_{,x}^2)-e^{-2\sigma}V(\phi)\right],
\label{T_xx}\ee
\be T^{\theta}_{\theta}=T^{\phi}_{\phi}=-e^{2\sigma}\left[\frac{1}{2}(-\phi_{,t}^2+\phi_{,x}^2)+e^{-2\sigma}V(\phi)\right],
\label{T_theta_theta}\ee
\be T_{uu}=\phi_{,u}^2,\ee
\be T_{vv}=\phi_{,v}^2,\ee
\be T=-e^{2\sigma}(-\phi_{,t}^2+\phi_{,x}^2)-4V(\phi).\label{Trace_T}\ee

The equations obtained in this Appendix can be used to derive the equations of motion as discussed in Sec.~\ref{sec:set_up_EoM}.

\section{\texorpdfstring{Spatial and temporal derivatives near the singularity curve for a Schwarzschild black hole}{Derivatives near the singularity curve for Schw. BH}\label{sec:appendix_B}}

In this Appendix, we derive the analytic expressions for the spatial and temporal derivatives near the singularity curve for a Schwarzschild black hole in Kruskal coordinates. Due to the similarity between Kruskal coordinates and double-null coordinates, these results can provide an intuitive understanding of the relation between the spatial and temporal derivatives near the singularity curve for the collapse in double-null coordinates.

For a Schwarzschild black hole in Kruskal coordinates, the expression for $r$ can be obtained from Eq.~(\ref{radius_schw_BH}):
\be \frac{r}{2m}=1+W(z),\label{r_W}\ee
where
\be z=\frac{x^2-t^2}{e},\nonumber\ee
and $W$ is the Lambert $W$ function defined by \cite{Corless}
\be Y=W(Y)e^{W(Y)}. \label{lambertW_definition} \ee
$Y$ can be a negative or a complex number. On the hypersurface of $r=\mbox{Const}$, $z=(x^2-t^2)/e=\mbox{Const}$. Then, in the two-dimensional
spacetime of $(t,x)$, the slope for the curve $r=\mbox{Const}$, $J$, can be expressed as
\be J\equiv\frac{dt}{dx}=\frac{x}{t}.\ee

The first- and second-order derivatives of $W$ are
\be \frac{dW}{dz}=\frac{W}{z(1+W)}, \hphantom{ddd} \mbox{for } z\neq \left\{0, -\frac{1}{e}\right\}, \label{W_z}\ee
\be \frac{d^{2}W}{dz^2}=-\frac{W^{2}(2+W)}{z^2(1+W)^{3}}, \hphantom{ddd} \mbox{for } z\neq \left\{0, -\frac{1}{e}\right\}. \label{W_zz}\ee
Consequently, with Eqs.~(\ref{r_W}), (\ref{W_z}), and (\ref{W_zz}), one can obtain the first- and second-order derivatives of $r$ with respect to $x$:
\be \frac{1}{2m}\cdot \frac{dr}{dx}=\frac{dW}{dz}\cdot \frac{2x}{e},\label{r_x}\ee
\be \frac{1}{2m}\cdot \frac{d^{2}r}{dx^2}=\frac{d^{2}W}{dz^2}\left(\frac{2x}{e}\right)^{2}+\frac{dW}{dz}\cdot \frac{2}{e}.\label{r_xx}\ee
Near the singularity curve, $z[=(x^2-t^2)/e]$ approaches $-1/e$, and $W$ asymptotes to $-1$. Consequently, the second-order derivative of $r$ with respect to $x$ can be approximated as follows:
\be \frac{1}{2m}\cdot \frac{d^{2}r}{dx^2}\approx -\frac{4x^2}{(1+W)^3} \approx \frac{d^{2}W}{dz^2}\left(\frac{2x}{e}\right)^{2}. \label{r_xx_v2}\ee
\vspace{1pt}

Similarly, one can obtain the first- and second-order derivatives of $r$ with respect to $t$ near the singularity curve:
\be \frac{1}{2m}\cdot \frac{dr}{dt}=-\frac{dW}{dz}\cdot \frac{2t}{e},\label{r_t}\ee
\be \frac{1}{2m}\cdot \frac{d^{2}r}{dt^2}\approx -\frac{4t^2}{(1+W)^3}\approx \frac{d^{2}W}{dz^2}\left(\frac{2t}{e}\right)^{2}. \label{r_tt}\ee
Therefore, with Eqs.~(\ref{r_x}) and (\ref{r_xx_v2})-(\ref{r_tt}), the ratios between the spatial and temporal derivatives can be expressed by the slope of the singularity curve, $J$:
\be \frac{\frac{dr}{dx}}{\frac{dr}{dt}}=-\frac{x}{t}=-J,\ee
\be \frac{\frac{d^{2}r}{dx^2}}{\frac{d^{2}r}{dt^2}}\approx\left(\frac{x}{t}\right)^2=J^2.\ee
As discussed in Sec.~\ref{sec:results_dynamics}, in spherical collapse in double-null coordinates, the ratios between the spatial and temporal derivatives are also defined by $J$.

\end{appendix}



\begin{thebibliography}{99}

\bibitem{Stelle}
K. S. Stelle,
{\it ``Renormalization of higher-derivative quantum gravity,''}
Phys. Rev. D {\bf 16}, 953 (1977).

\bibitem{Starobinsky_1}
A. A. Starobinsky,
{\it ``A new type of isotropic cosmological models without singularity,''}
Phys. Lett. {\bf 91}B, 99 (1980).

\bibitem{Biswas_1}
T. Biswas, E. Gerwick, T. Koivisto, and A. Mazumdar,
{\it ``Towards singularity and ghost free theories of gravity,''}
Phys. Rev. Lett. {\bf 108}, 031101 (2012).
[\arXiv[gr-qc]{1110.5249}]

\bibitem{Biswas_2}
T. Biswas, A. Conroy, A. S. Koshelev, and A. Mazumdar,
{\it ``Generalized ghost-free quadratic curvature gravity,''}
Classical Quantum Gravity {\bf 31}, 015022 (2014).
[\arXiv[hep-th]{1308.2319}]

\bibitem {Brans_1961}
C. H. Brans and R. H. Dicke,
{\it ``Mach's principle and a relativistic theory of gravitation,''}
Phys. Rev. {\bf 124}, 925 (1961).

\bibitem{Milgrom}
M. Milgrom,
{\it ``A modification of the Newtonian dynamics as a possible alternative to the hidden mass hypothesis,''}
Astrophys. J. {\bf 270}, 365 (1983).

\bibitem{Damour}
T. Damour and G. Esposito-Farese,
{\it ``Tensor-multi-scalar theories of gravitation,''}
Classical Quantum Gravity {\bf 9}, 2093  (1992).

\bibitem{Carroll}
S. M. Carroll, V. Duvvuri, M. Trodden, and M. S. Turner,
{\it ``Is Cosmic Speed-Up Due to New Gravitational Physics?''}
Phys. Rev. D \textbf{70}, 043528 (2004).
[\arXiv{astro-ph/0306438}]

\bibitem{Hu_0705}
W. Hu and I. Sawicki,
{\it ``Models of f(R) Cosmic Acceleration that Evade Solar-System Tests,''}
Phys. Rev. D {\bf 76}, 064004 (2007).
[\arXiv[astro-ph]{0705.1158}]

\bibitem{Starobinsky_2}
A. A. Starobinsky,
{\it ``Disappearing cosmological constant in f(R) gravity,''}
JETP Lett., \textbf{86}, 157 (2007).
[\arXiv[astro-ph]{0706.2041}]

\bibitem{Sotiriou_1}
T. P. Sotiriou and V. Faraoni,
{\it ``f(R) Theories Of Gravity,''}
Rev. Mod. Phys. {\bf 82}, 451 (2010).
[\arXiv[gr-qc]{0805.1726}]

\bibitem{Tsujikawa1}
A. D. Felice and S. Tsujikawa,
{\it ``f (R) Theories,''}
Living Rev. Relativity \textbf{13}, 3 (2010).
[\arXiv[gr-qc]{1002.4928}]

\bibitem{Berger_2002}
B. K. Berger,
{\it ``Numerical Approaches to Spacetime Singularities,''}
Living Rev. Relativity {\bf 5}, 1 (2002).
[\arXiv{gr-qc/0201056}]

\bibitem{Joshi_2007}
P. S. Joshi,
{\it Gravitational Collapse and Spacetime Singularities}
(Cambridge University Press, Cambridge, UK, 2007).

\bibitem{Henneaux}
M. Henneaux, D. Persson, and P. Spindel,
{\it ``Spacelike Singularities and Hidden Symmetries of Gravity,''}
Living Rev. Relativity {\bf 11}, 1 (2008).
[\arXiv[hep-th]{0710.1818}]

\bibitem{Kamenshchik}
A. Yu. Kamenshchik,
{\it ``The problem of singularities and chaos in cosmology,''}
Phys. -Usp. {\bf 53}, 301 (2010).
[\arXiv[gr-qc]{1006.2725}]

\bibitem{Joshi_2011}
P. S. Joshi and D. Malafarina,
{\it ``Recent developments in gravitational collapse and spacetime singularities,''}
Int. J. Mod. Phys. D{\bf 20}, 2641 (2011).
[\arXiv[gr-qc]{1201.3660}]

\bibitem{Belinski_1404}
V. A. Belinski,
{\it ``On the cosmological singularity,''}
\arXiv[gr-qc]{1404.3864}

\bibitem{Wheeler}
R. Ruffini and J. A. Wheeler,
{\it ``Introducing the black hole,''}
Phys. Today {\bf 24}, No. 1, 30 (1971).

\bibitem{Hawking}
S. W. Hawking,
{\it ``Black holes in the Brans-Dicke theory of gravitation,''}
Comm. Math. Phys. {\bf 25}, 167 (1972).

\bibitem{Bekenstein}
J. D. Bekenstein,
{\it ``Novel \lq no-scalar-hair\rq~theorem for black holes,''}
Phys. Rev. D {\bf 51}, R6608 (1995).

\bibitem{Sotiriou_2}
T. P. Sotiriou and V. Faraoni,
{\it ``Black holes in scalar-tensor gravity,''}
Phys. Rev. Lett. {\bf 108}, 081103 (2012).
[\arXiv[gr-qc]{1109.6324}]

\bibitem{Oppenheimer}
J. R. Oppenheimer and H. Snyder,
{\it ``On Continued Gravitational Contraction,''}
Phys. Rev. {\bf 56}, 455 (1939).

\bibitem{Lemaitre}
G. Lema\^{\i}tre,
{\it ``The Expanding Universe,''}
Annales Soc. Sci. Brux. Ser. I Sci. Math. Astron. Phys. A {\bf 53}, 51 (1933).
Reprint: Gen. Rel. Grav. {\bf 29} 641 (1997).

\bibitem{Tolman}
R. C. Tolman,
{\it ``Effect of Inhomogeneity on Cosmological Models,''}
Proc. Natl. Acad. Sci. U.S.A. {\bf 20}, 169 (1934).
Reprint: Gen. Relativ. Gravit. {\bf 29}, 935 (1997).

\bibitem{Bondi}
H. Bondi,
{\it ``Spherically symmetrical models in general relativity,''}
Mon. Not. R. Astron. Soc. {\bf 107}, 410 (1947).

\bibitem{Shibata}
M. Shibata, K. Nakao, and T. Nakamura,
{\it ``Scalar-type gravitational wave emission from gravitational collapse in Brans-Dicke theory: Detectability by a laser interferometer,''}
Phys. Rev. D {\bf 50}, 7304 (1994).

\bibitem{Scheel_1}
M. A. Scheel, S. L. Shapiro, and S. A. Teukolsky,
{\it ``Collapse to Black Holes in Brans-Dicke Theory: I. Horizon Boundary Conditions for Dynamical Spacetimes,''}
Phys. Rev. D {\bf 51}, 4208 (1995).
[\arXiv{gr-qc/9411025}]

\bibitem{Scheel_2}
M. A. Scheel, S. L. Shapiro, and S. A. Teukolsky,
{\it ``Collapse to Black Holes in Brans-Dicke Theory: II. Comparison with General Relativity,''}
Phys. Rev. D {\bf 51}, 4236 (1995).
[\arXiv{gr-qc/9411026}]

\bibitem{Hertog}
T. Hertog,
{\it ``Towards a Novel no-hair Theorem for Black Holes,''}
Phys. Rev. D {\bf 74}, 084008 (2006).
[\arXiv{gr-qc/0608075}]

\bibitem{Berti}
E. Berti, V. Cardoso, L. Gualtieri, M. Horbatsch, and U. Sperhake,
{\it ``Numerical simulations of single and binary black holes in scalar-tensor theories: circumventing the no-hair theorem,''}
Phys. Rev. D {\bf 87}, 124020 (2013).
[\arXiv[gr-qc]{1304.2836}]

\bibitem{Cembranos}
J. A. R. Cembranos, A. de la Cruz-Dombriz, and B. M. Nunez,
{\it ``Gravitational collapse in f(R) theories,''}
J. Cosmol. Astropart. Phys. {\bf 04} (2012) 021.
[\arXiv[gr-qc]{1201.1289}]

\bibitem{Senovilla}
J. M. M. Senovilla,
{\it ``Junction conditions for f(R) gravity and their consequences,''}
Phys. Rev. D {\bf 88}, 064015 (2013).
[\arXiv[gr-qc]{1303.1408}]

\bibitem{Hwang_1110}
D.-i. Hwang, B.-H. Lee, and D.-h. Yeom,
{\it ``Mass inflation in f(R) gravity: A conjecture on the resolution of the mass inflation singularity,''}
J. Cosmol. Astropart. Phys. {\bf 12} (2011) 006.
[\arXiv[gr-qc]{1110.0928}]

\bibitem{Kopp}
M. Kopp, S. A. Appleby, I. Achitouv, and J. Weller,
{\it ``Spherical collapse and halo mass function in f(R) theories,''}
Phys. Rev. D {\bf 88}, 084015 (2013).
[\arXiv[astro-ph.CO]{1306.3233}]

\bibitem{Barreira}
A. Barreira, B. Li, C. Baugh, and S. Pascoli,
{\it ``Spherical collapse in Galileon gravity: fifth force solutions, halo mass function and halo bias,''}
J. Cosmol. Astropart. Phys. {\bf 11} (2013) 056.
[\arXiv[astro-ph.CO]{1308.3699}]

\bibitem{Belinskii}
V. A. Belinskii, I. M. Kalathnikov, and E. M. Lifshitz,
{\it ``Oscillatory Approach to a Singular Point in the Relativistic Cosmology,''}
Adv. Phys. {\bf 19}, 525 (1970).

\bibitem{Landau}
L. D. Landau and E. M. Lifshitz,
{\it The Classical Theory of Fields}, Course of Theoretical Physics Series Vol.2
(Pergamon Press, Oxford, UK 1971), 4th ed.

\bibitem{Kasner}
E. Kasner,
{\it ``Geometrical theorems on Einstein's cosmological equations,''}
Am. J. Math, {\bf 43}, 217 (1921).

\bibitem{Wainwright}
J. Wainwright and A. Krasinski,
{\it ``Republication of: Geometrical theorems on Einstein's cosmological equations (By E. Kasner),''}
Gen. Relativ. Gravit. {\bf 40}, 865 (2008).

\bibitem{Berger}
B. K. Berger, D. Garfinkle, J. Isenberg, V. Moncrief, and M. Weaver,
{\it ``The Singularity in Generic Gravitational Collapse Is Spacelike, Local, and Oscillatory,''}
Mod. Phys. Lett. A \textbf{13}, 1565 (1998).
[\arXiv{gr-qc/9805063}]

\bibitem{Garfinkel_1}
D. Garfinkle,
{\it ``Numerical Simulations of Generic Singularities,''}
Phys. Rev. Lett. \textbf{93}, 161101 (2004).
[\arXiv{gr-qc/0312117}]

\bibitem{Garfinkel_2}
R. Saotome, R. Akhoury, and D. Garfinkle,
{\it ``Examining Gravitational Collapse With Test Scalar Fields,''}
Classical Quantum Gravity \textbf{27}, 165019 (2010).
[\arXiv[gr-qc]{1004.3569}]

\bibitem{Ashtekar}
A. Ashtekar, A. Henderson, and D. Sloan,
{\it ``A Hamiltonian Formulation of the BKL Conjecture,''}
Phys. Rev. D \textbf{83}, 084024 (2011).
[\arXiv[gr-qc]{1102.3474}]

\bibitem{Harada}
T. Harada, T. Chiba, K.-I. Nakao, and T. Nakamura,
{\it ``Scalar gravitational wave from Oppenheimer-Snyder collapse in scalar-tensor theories of gravity,''}
Phys. Rev. D {\bf 55}, 2024 (1997).
[\arXiv{gr-qc/9611031}]

\bibitem{Sotani}
H. Sotani,
{\it ``Scalar gravitational waves from relativistic stars in scalar-tensor gravity,''}
Phys. Rev. D {\bf 89} 064031 (2014).
[\arXiv[astro-ph]{1402.5699}]

\bibitem{Hwang_2010}
D.-i. Hwang and D.-h. Yeom,
{\it ``Responses of the Brans-Dicke field due to gravitational collapses,''}
Classical Quantum Gravity {\bf 27}, 205002 (2010).
[\arXiv[gr-qc]{1002.4246}]

\bibitem{Frolov_0504}
A. V. Frolov, K. R. Kristjansson, L. Thorlacius,
{\it ``Semi-classical geometry of charged black holes,''}
Phys. Rev. D {\bf 72}, 021501 (2005).
[\arXiv{hep-th/0504073}]

\bibitem{Frolov_0604}
A. V. Frolov, K. R. Kristjansson, L. Thorlacius,
{\it ``Global geometry of two-dimensional charged black holes,''}
Phys. Rev. D {\bf 73}, 124036 (2006).
[\arXiv{hep-th/0604041}]

\bibitem{Borkowska}
A. Borkowska, M. Rogatko, and R. Moderski,
{\it ``Collapse of Charged Scalar Field in Dilaton Gravity,''}
Phys. Rev. D {\bf 83}, 084007 (2011).
[\arXiv[gr-qc]{1103.4808}]

\bibitem{Cao}
Z. Cao, P. Galaviz, and L.-F. Li,
{\it ``Binary black hole mergers in f(R) theory,''}
Phys. Rev. D {\bf 87}, 104029 (2013).

\bibitem{Nunez}
A. Nunez and S. Solganik,
{\it ``The content of f(R) gravity,''}
\arXiv{hep-th/0403159}

\bibitem{Dolgov}
A. D. Dolgov and M. Kawasaki,
{\it ``Can modified gravity explain accelerated cosmic expansion?''}
Phys. Lett. B {\bf 573}, 1 (2003).
[\arXiv{astro-ph/0307285}]

\bibitem {Amendola}
L. Amendola, R. Gannouji, D. Polarski, and S. Tsujikawa,
{\it ``Conditions for the cosmological viability of f(R) dark energy models,''}
Phys. Rev. D {\bf 75}, 083504 (2007).
[\arXiv{gr-qc/0612180}]

\bibitem{Guo_1305}
J.-Q. Guo and A. V. Frolov,
{\it ``Cosmological dynamics in f(R) gravity,''}
Phys. Rev. D {\bf 88}, 124036 (2013).
[\arXiv[astro-ph.CO]{1305.7290}]

\bibitem{Justin1}
J. Khoury and A. Weltman,
{\it ``Chameleon Fields: Awaiting Surprises for Tests of Gravity in Space,''}
Phys. Rev. Lett. {\bf 93}, 171104 (2004).
[\arXiv{astro-ph/0309300}]

\bibitem{Justin2}
J. Khoury and A. Weltman,
{\it ``Chameleon Cosmology,''}
Phys. Rev. D {\bf 69}, 044026 (2004).
[\arXiv{astro-ph/0309411}]

\bibitem{Chiba}
T. Chiba, T. L. Smith, and A. L. Erickcek,
{\it ``Solar System constraints to general f(R) gravity,''}
Phys. Rev. D {\bf 75}, 124014 (2007).
[\arXiv{astro-ph/0611867}]

\bibitem{Tamaki_0808}
T. Tamaki and S. Tsujikawa,
{\it ``Revisiting chameleon gravity - thin-shells and no-shells with appropriate boundary conditions,''}
Phys. Rev. D {\bf 78}, 084028 (2008).
[\arXiv[gr-qc]{0808.2284}]

\bibitem{Tsujikawa_0901}
S. Tsujikawa, T. Tamaki, and R. Tavakol,
{\it ``Chameleon scalar fields in relativistic gravitational backgrounds,''}
J. Cosmol. Astropart. Phys. {\bf 05} (2009) 020.
[\arXiv[gr-qc]{0901.3226}]

\bibitem{Guo_1306}
J.-Q. Guo,
{\it ``Solar system tests of f(R) gravity,''}
Int. J. Mod. Phys. D{\bf 23}, 1450036 (2014).
[\arXiv[astro-ph.CO]{1306.1853}]

\bibitem{Christodoulou}
D. Christodoulou,
{\it ``Bounded Variation Solutions of the Spherically Symmetric Einstein-Scalar Field Equations,''}
Commun. Pure Appl. Math. \textbf{46}, 1131 (1993).

\bibitem{Frolov_2004}
A. V. Frolov,
{\it ``Is It Really Naked? On Cosmic Censorship in String Theory,''}
Phys. Rev. D {\bf 70}, 104023 (2004).
[\arXiv{hep-th/0409117}]

\bibitem{Sorkin}
E. Sorkin and T. Piran,
{\it ``Effects of Pair Creation on Charged Gravitational Collapse,''}
Phys. Rev. D {\bf 63}, 084006 (2001).
[\arXiv{gr-qc/0009095}]

\bibitem{Hwang_1105}
D.-i. Hwang, H. Kim, and D.-h. Yeom,
{\it ``Dynamical formation and evolution of (2+1)-dimensional charged black holes,''}
Classical Quantum Gravity {\bf 29}, 055003 (2012).
[\arXiv[gr-qc]{1105.1371}]

\bibitem{Golod}
S. Golod and T. Piran,
{\it ``Choptuik's Critical Phenomenon in Einstein-Gauss-Bonnet Gravity,''}
Phys. Rev. D {\bf 85}, 104015 (2012).
[\arXiv[gr-qc]{1201.6384}]

\bibitem{Pretorius}
F. Pretorius,
{\it ``Numerical Relativity Using a Generalized Harmonic Decomposition,''}
Classical Quantum Gravity {\bf 22}, 425 (2005).
[\arXiv{gr-qc/0407110}]

\bibitem{Baumgarte}
T. W. Baumgarte and S. L. Shapiro,
{\it Numerical relativity: Solving Einstein's Equations on the Computer}
(Cambridge University Press, Cambridge, UK, 2010).

\bibitem{Csizmadia}
P. Csizmadia and I. Racz,
{\it ``Gravitational collapse and topology change in spherically symmetric dynamical systems,''}
Classical Quantum Gravity {\bf 27}, 015001  (2010).
[\arXiv[gr-qc]{0911.2373}]

\bibitem{Garfinkel_3}
D. Garfinkle,
{\it ``Choptuik scaling in null coordinates,''}
Phys. Rev. D {\bf 51}, 5558 (1995).
[\arXiv{gr-qc/9412008}]

\bibitem{Nariai}
H. Nariai,
{\it ``Hamiltonian approach to the dynamics of expanding homogeneous universes in the Brans-Dicke cosmology,''}
Prog. Theor. Phys. {\bf 47}, 1824 (1972).

\bibitem{Belinskii_2}
V. A. Belinskii and I. M. Khalatnikov,
{\it ``Effect of scalar and vector fields on the nature of the cosmological singularity,''}
Zh. Eksp. Teor. Fiz. {\bf 63}, 1121 (1972)
[Sov. Phys. JETP {\bf 36}, 591 (1973)].

\bibitem{Wald_1984}
R. M. Wald,
{\it General Relativity}
(The University of Chicago Press, Chicago, U.S.A., 1984).

\bibitem{Frolov_2}
A. V. Frolov,
{\it ``A Singularity Problem with f(R) Dark Energy,''}
Phys. Rev. Lett. \textbf{101}, 061103 (2008).
[\arXiv[astro-ph]{0803.2500}]

\bibitem{Corless}
R. M. Corless, G. H. Gonnet, D. E. G. Hare, D. J. Jeffrey, and D. E. Knuth,
{\it ``On the Lambert W function,''}
Adv. Comput. Math. {\bf 5}, 329 (1996).

\end{thebibliography}
\end{document}